\documentclass{vgtc}                          

\newif\ifwarntodo 
\warntodotrue

\newif\ifshowtodo
\showtodotrue

\usepackage{times}                     

\usepackage{tabu}                      
\usepackage{booktabs}                  
\usepackage{lipsum}                    
\usepackage{mwe}                       

\usepackage{mathptmx}                  
\graphicspath{{pdfimages/}, {images/}} 
\usepackage{subcaption}   

\usepackage{booktabs}
\usepackage{multirow}
\usepackage{microtype}

\usepackage[svgnames]{xcolor}

\usepackage{makecell}   

\usepackage{tikz}
\usetikzlibrary{patterns,shapes.arrows}
\usetikzlibrary{positioning}
\usetikzlibrary{arrows.meta}

\usetikzlibrary{external}
\tikzset{external/only named=true}

\usepackage{tikzscale}

\usepackage{pgfplots}
\usepackage{pgfplotstable}
\usepgfplotslibrary{groupplots}
\pgfplotsset{compat=newest}

\usepackage{amsmath}
\usepackage{amsbsy}
\usepackage{amsfonts}
\usepackage{mathtools}
\usepackage{amssymb, amsthm}  
\usepackage{nicefrac}

\usepackage{cite}                      

\onlineid{1028}

\vgtccategory{Research}

\vgtcinsertpkg

\title{A Comparative Study of Different Edit Distance-Based Methods\\
for Feature Tracking using Merge Trees\\on Time-Varying Scalar Fields}

\author{Son Le Thanh\thanks{e-mail: sonlt@kth.se} %
\and Tino Weinkauf\thanks{e-mail: weinkauf@kth.se} %
}
\affiliation{\scriptsize KTH Royal Institute of Technology\\Stockholm, Sweden}

\abstract{
    Feature tracking in time-varying scalar fields is a fundamental task in scientific computing.
    Topological descriptors, which 
    \changed{summarize} 
    important features of data, 
    \changed{have}
    proved to be viable tools
    to facilitate this task.  
    The merge tree is a topological descriptor 
    that captures the connectivity behaviors of the sub- or superlevel sets 
    of a scalar field.
    Edit distances 
    between merge trees play
    a vital role in effective temporal data tracking.
    Existing methods to compute them
    fall into two main classes, namely
		whether they are dependent or independent of the branch decomposition.
    These two classes represent the most prominent approaches for 
    producing tracking results.
    In this paper,
    \changed{we compare 
    four different merge tree edit distance-based methods 
    for feature tracking}. 
    We demonstrate that these methods yield distinct results with both analytical
    and real-world data sets.
    Furthermore, we investigate how these results vary
    and identify the factors that influence them.  
    Our experiments reveal significant differences in tracked features over time, 
    even among those produced by techniques within the same category.
} 


\begin{document}



\newcommand{\RRR}{{\mathrm I\! \mbox{R} }}
\newcommand{\EEE}{{\mathrm I\! \mbox{E} }}
\newcommand{\xx}{{\mathbf x}}
\newcommand{\yy}{{\mathbf y}}
\newcommand{\zz}{{\mathbf z}}
\newcommand{\dd}{{\mathbf d}}
\newcommand{\hh}{{\mathbf h}}
\newcommand{\ttt}{{\mathbf t}}
\newcommand{\jj}{{\mathbf j}}
\newcommand{\pp}{{\mathbf p}}
\newcommand{\qq}{{\mathbf q}}
\newcommand{\aaa}{{\mathbf a}}
\newcommand{\ssss}{{\mathbf s}}
\newcommand{\bb}{{\mathbf b}}
\newcommand{\ee}{{\mathbf e}}
\newcommand{\cc}{{\mathbf c}}
\newcommand{\nn}{{\mathbf n}}
\newcommand{\mm}{{\mathbf m}}
\newcommand{\kk}{{\mathbf k}}
\newcommand{\rr}{{\mathbf r}}
\newcommand{\uu}{{\mathbf u}}
\newcommand{\vv}{{\mathbf v}}
\newcommand{\ff}{{\mathbf f}}
\newcommand{\fff}{{\mathbf f}}
\newcommand{\JJ}{{\mathbf J}}
\newcommand{\DD}{{\mathbf D}}
\newcommand{\BB}{{\mathbf B}}
\newcommand{\CC}{{\mathbf C}}
\newcommand{\EE}{{\mathbf E}}
\newcommand{\PP}{{\mathbf P}}
\newcommand{\MM}{{\mathbf M}}
\newcommand{\LL}{{\mathbf L}}
\newcommand{\QQ}{{\mathbf Q}}
\newcommand{\FF}{{\mathbf F}}
\newcommand{\TT}{{\mathbf T}}
\newcommand{\RR}{{\mathbf R}}
\newcommand{\SSS}{{\mathbf S}}
\newcommand{\Sa}{{\mathbf {Sa}}}
\newcommand{\ZZ}{Z}
\newcommand{\cs}{{\dot c}}
\newcommand{\sss}{{\dot s}}
\newcommand{\ps}{{\dot p}}
\newcommand{\pg}{{\dot g}}
\newcommand{\ww}{{\mathbf w}}
\newcommand{\xxs}{\dot{\mathbf x}}
\newcommand{\vvs}{\dot{\mathbf v}}
\newcommand{\us}{{\dot u}}
\newcommand{\vs}{{\dot v}}
\newcommand{\ws}{{\dot w}}
\newcommand{\fn}{{\mathbf 0}}

\newcommand{\ii}{{\mathbf i}}
\newcommand{\Dt}{{\widetilde D}}
\newcommand{\fft}{{\widetilde \ff}}
\newcommand{\kkt}{{\widetilde \kk}}
\newcommand{\ggt}{{\widetilde \ggg}}
\newcommand{\hht}{{\widetilde \hh}}
\newcommand{\vvt}{{\widetilde \vv}}
\newcommand{\wwt}{{\widetilde \ww}}

\newcommand{\dbar}[1]{\bar{\bar{#1}}}
\newcommand{\qqqq}{\dbar{\qq}}

\newcommand{\eet}{\dbar{\ee}_{t}}

\newcommand{\eetau}{\dbar{\ee}_{\tau}}
%

\newcommand{\mNN}{\mathcal N}
\newcommand{\mEE}{\mathcal E}
\newcommand{\mSS}{\mathcal S}
\newcommand{\mRR}{\mathcal R}
\newcommand{\mPP}{\mathcal P}
\newcommand{\mGG}{\mathcal G}
\newcommand{\mHH}{\mathcal H}
\newcommand{\mVV}{\mathcal V}
\newcommand{\mTT}{\mathcal T}
\newcommand{\mLL}{\mathcal L}
\newcommand{\mBB}{\mathcal B}

\newcommand{\UF}{\texttt{UF}\ }

\newtheorem{theorem}{Theorem}
\newtheorem{lemma}{Lemma}
\newtheorem{corollary}{Corollary}

\newcommand{\dataset}[1]{\texttt{#1}}

\newcommand{\magn}[1]{|#1|}
\newcommand{\cardinality}[1]{|#1|}
\newcommand{\norm}[1]{\left\lVert#1\right\rVert}

\newcommand{\leaves}[1]{\ell(#1)}
\newcommand{\treeroot}[1]{\text{root}(#1)}
\newcommand{\inner}[1]{\text{inner}(#1)}

\newcommand{\pathvertices}[1]{\text{vert}(#1)}
\newcommand{\edges}[1]{\textit{e}(#1)}
\newcommand{\bdt}[1]{\text{BDT}(#1)}

\newcommand{\idx}[1]{\text{idx}(#1)}
\newcommand{\len}[1]{\text{len}(#1)}
\newcommand{\pdiff}[1]{\textit{d}(#1)}

%
%




\ifwarntodo
	\newcommand{\warn}{\PackageWarning{}{Unprocessed note}}
\else
	\newcommand{\warn}{}
\fi

\newcommand{\todo}[1]{\warn%
\ifshowtodo%
\textbf{\small\textcolor{red}{TODO: #1}}%
\fi%
}

\newcommand{\annoauthor}[3]{\warn%
\ifshowtodo%
\textbf{\small\textcolor[rgb]{#1}{#2: \emph{#3}}}%
\fi%
}

\newcommand{\son}[1]{\annoauthor{0.2,0.2,0.8}{Son}{#1}}
\newcommand{\tino}[1]{\annoauthor{0.8,0.2,0.2}{Tino}{#1}}

\newcommand{\changedCamera}[1]{#1}
\newcommand{\changed}[1]{#1}
\newcommand{\allchanged}{}
\newcommand{\alladded}{}
\newcommand{\allsame}{}
\newcommand{\added}[1]{#1}

\newlength{\lengthgoodgap}
\addtolength{\lengthgoodgap}{12pt}
\newcommand{\goodgap}{\hspace{\lengthgoodgap}}

\newcommand{\goodnewline}{\\[0.5\lengthgoodgap]}

\newlength{\twopicwidth}
\addtolength{\twopicwidth}{0.5\textwidth}
\addtolength{\twopicwidth}{-0.5\lengthgoodgap}

\newlength{\threepicwidth}
\addtolength{\threepicwidth}{0.333333\textwidth}
\addtolength{\threepicwidth}{-0.666666\lengthgoodgap}

\newlength{\fourpicwidth}
\addtolength{\fourpicwidth}{0.25\textwidth}
\addtolength{\fourpicwidth}{-0.75\lengthgoodgap}

\newlength{\fivepicwidth}
\addtolength{\fivepicwidth}{0.20\textwidth}
\addtolength{\fivepicwidth}{-0.80\lengthgoodgap}

\newlength{\sixpicwidth}
\addtolength{\sixpicwidth}{0.166666666666666667\textwidth}
\addtolength{\sixpicwidth}{-0.833333333333333333\lengthgoodgap}

\newlength{\ltwopicwidth}
\addtolength{\ltwopicwidth}{0.5\linewidth}
\addtolength{\ltwopicwidth}{-0.5\lengthgoodgap}

\newlength{\lthreepicwidth}
\addtolength{\lthreepicwidth}{0.333333\linewidth}
\addtolength{\lthreepicwidth}{-0.666666\lengthgoodgap}

\newlength{\lfourpicwidth}
\addtolength{\lfourpicwidth}{0.25\linewidth}
\addtolength{\lfourpicwidth}{-0.75\lengthgoodgap}

\hfuzz=3pt

\def\chapterautorefname{Chapter}
\def\sectionautorefname{Section}
\def\subsectionautorefname{Section}
\def\figureautorefname{Figure}
\def\subfigureautorefname{Figure}
\def\tableautorefname{Table}
\def\equationautorefname~#1\null{(#1)\null}

\newcommand{\Autoref}[1]{%
  \begingroup%
  \def\chapterautorefname{Chapter}%
  \def\sectionautorefname{Section}%
  \def\subsectionautorefname{Section}%
  \def\figureautorefname{Figure}%
	\def\subfigureautorefname{Figure}
  \def\tableautorefname{Table}%
  \def\equationautorefname~##1\null{Equation~(##1)\null}%
  \autoref{#1}%
  \endgroup%
}

\newcommand{\SupplementalMaterialHeadWithNames}[1]{%
\begin{center}
Supplemental Material -- #1
\end{center}

\vspace{-1.5\baselineskip}
\section*{\centering Temporal Merge Tree Maps:\\A Topology-Based Static Visualization for Temporal Scalar Data}
\subsection*{\centering Wiebke Köpp and Tino Weinkauf}
}

\newcommand{\resetlength}[2]{\ifx#1\undefined \newlength{#1} \fi \setlength{#1}{#2} }

%
%





\maketitle

\section{Introduction} 
\label{sec-Intro}

Tracking features is a classic task in feature-based data analysis and visualization.
Tracked features reveal important spatio-temporal characteristics of a data set.
For example,
tracking cyclones in meteorological data \cite{neu13}
is important to understand water transport and other atmospheric phenomena.

Many different methods to feature tracking have been introduced in the past.
This includes tracking of
critical points \cite{theisel03b, weinkauf10a, tricoche02, garth04b},
vortex core lines \cite{bauer02, theisel05b, weinkauf07c, Guenther15SciVis},
or regions defined by thresholding \cite{silverwang1997, silverwang1998}.

In the last decade,
several methods were proposed
to track merge trees in time-dependent scalar fields.
A merge tree
is a hierarchical description
of the evolving connectivity
of the sub- / super-levelsets
when considering an increasing/decreasing
threshold for a scalar field.
Most methods extract the merge tree in each time step,
possibly apply a simplification step to reduce the topological complexity,
and then apply a secondary algorithm
for matching the (simplified) trees of adjacent time steps.
\changed{Recently, edit distance-based methods,}
namely the constrained edit distance \cite{sridharamurthy2018edit},
the Wasserstein distance \cite{pont22},
the branch mapping \cite{wetzels2022branch}, and 
path mapping \cite{wetzels2022path} distances,
\changed{have demonstrated their effectiveness for 
feature tracking.}

Little to no work has been done so far
on comparing the outcomes of these algorithms.
In fact,
assessing the quality of the tracking results
is not straightforward.
In this paper,
we present methods for comparing 
\changed{results from merge tree edit distance-based
tracking approaches, considering} 
their individual characteristics
as well as
their pairwise agreement on feature matching.
\changed{We exclude more advanced and hybrid tracking approaches
and left these works for further investigations.}
We hope to contribute
to a discussion in the community
on how to assess merge tree tracking results
in the absence of a ground truth.

We review the basic math around merge trees in \Cref{sec-Background}.
The four compared tracking methods
are described in \Cref{sec-TrackingMethods}.
The majority of the paper
is devoted to the comparison of these methods
in \Cref{sec-experiments}.
Discussion and conclusions follow
in \Cref{sec-discussion} and \Cref{sec-conclusion}.

\section{Preliminaries}
\label{sec-Background}
In this section, we discuss the core concepts 
of merge tree, its derivation branch decomposition tree,
and the definition for the distance between two trees.  

\subsection{Merge Trees}
\label{sec-MergeTrees}
We state the formal definition of merge trees 
and follow with a less formal interpretation for them.
Consider a scalar field $f: \MM \rightarrow \RR$ 
defined on a
%
%
manifold $\MM$. 
Let $f^{-}(-\infty, a] = \{\pp \in \MM | f(\pp) \leq a \}$ 
be a \emph{sublevel set} of $f$ for $a\in \RR$.
Consider an equivalence relation
that describes connected components of isocontours:
two points $\pp_1, \pp_2 \in \MM$
are called equivalent $\pp_1 \sim \pp_2$, iff 
$f(\pp_1) = f(\pp_2)$
and $\pp_1$ and $\pp_2$ belong to the same connected component
of $f^{-}(-\infty, a]$. 
The quotient set $\TT^{-}(\MM, f) = \MM / \sim$, 
obtained by identifying or ``gluing'' points in $\MM$ that are equivalent 
under the defined relation $\sim$, is called a \emph{join tree}.
Analogously, consider the set $\TT^{+}(\MM, f) = \MM / \sim$
where the equivalence relation is defined using the \emph{superlevel set}
$f^{+}[a, \infty) = \{ \pp \in \MM | f(\pp) \geq a\}$ for $a \in \RR$. 
The quotient set $\TT^{+}(\MM, f)$ is a \emph{split tree}. 
Split trees and join trees collectively
refer to as \emph{merge trees}, denoted by $\TT(\MM, f)$.
We can drop either the symbol $\MM$ or both of the symbols and
refer to the merge tree as $\TT$ if the manifold $\MM$
and $f$ are understood or they are not necessary for the discussion.

The merge tree $\TT(f)$ describes the connectivity behavior 
of the sub- or superlevel set of the scalar field $f$ 
using a tree structure.
The \emph{nodes} of $\TT(f)$ are the \emph{critical points}
of $f$.
For the case of the join tree, the \emph{leaves} of this tree corresponds 
to the \emph{local minima} of the scalar field. 
Each minimum gives rise to a 
\changed{\emph{sub-level set component}}.
As the isovalues are swept from $-\infty$ to $\infty$, 
the 
\changed{components} 
join at \emph{saddle points} which are represented 
by \emph{inner nodes} connecting to the leaves of the tree.
The sweeping process stops when all 
\changed{components} 
are connected 
at the \emph{global maximum}, i.e., the \emph{root} of the tree.
Conversely, for a split tree,
the leaves and the root are the \emph{local maxima} and \emph{global minimum},
respectively.
An \emph{edge} of a merge tree $\TT(f)$ connecting two critical points
can be weighted by the area or the volume of the region it occupies,
or the function difference between the nodes.
The later weighting method is closely related to the concept of \emph{persistence} \cite{edelsbrunner02}.

\subsection{Branch Decomposition Trees}
\label{sec-BDT}
Consider a tree $\TT = (V, E)$,
where $V = \{\vv_1,\dots,\vv_k \}$ and
$E = \{(\vv_1, \vv_2) | \vv_1, \vv_2 \in V \} $
are the set of the nodes and edges of $\TT$, respectively.
Let $\leaves{\TT} \subset V$ be the set of leaves of $\TT$.
A \emph{path} $p$ in this tree is a sequence of nodes
$p = (\vv_i)_{i = 1}^k \subset V^k$ such that
$(\vv_i, \vv_{i + 1}) \in E$ for all $1 \leq i \leq k - 1$.
For a path $p$, let $\edges{p} = \{(\vv_i, \vv_{i + 1})| (\vv_i, \vv_{i+1}) \in E \}$
be the set of the edges contained in this path.
A \emph{branch} $b = (\vv_i)_{i = 1}^k$ of $\TT$ is a path  
that terminates at a leaf, i.e., $\vv_k \in \leaves{\TT}$.
Suppose that $b_1 = (\vv_i)_{i = 1}^{k_1}$ and 
$b_2 = (\uu_j)_{j = 1}^{k_2}$ are two branches of $\TT$.
If for some $1 < i < k_1$, $\uu_1 = \vv_i$,
then $b_1$ is called a \emph{parent branch} of $b_2$, 
or $b_2$ is a \emph{child branch} of $b_1$, denoted by $b_1 \geq b_2$.

The set of branches of $B(\TT) = \{b_1,\dots,b_m \}$
such that
$\edges{b_i} \cap \edges{b_j} = \varnothing$ and  
$E = \cup_{i=1}^m \edges{b_i}$ is called a \emph{branch decomposition}
of $\TT$.
%
%
For a branch decomposition $B(\TT)$,
consider the tree whose vertex set is $B(\TT)$ and 
the edge set is $\{(b_p, b_c) | b_p \geq b_c \}$.  
This tree is called a \emph{branch decomposition tree} of $\TT$.
For a merge tree $\TT(f)$, denote the branch decomposition tree by $\bdt{f}$.
Note that the branch decomposition tree is not unique.
A merge tree $\TT$ can admit multiple branch decomposition trees.


\subsection{Edit Distance}
\label{section-editdistance}


Edit distance between merge trees plays an important role
in many recent proposals on feature tracking using merge trees.
We briefly review the definition of edit distance between two trees.
The original tree edit distance between two rooted ordered labeled trees
was introduced by Tai \cite{tai1979}.
For a rooted ordered labeled tree $\TT$, let $\Sigma$ be the set of 
its labels and $\varnothing \notin \Sigma$ the empty character. 
For two labels $a,b \in \Sigma$, the edit operations include
\begin{itemize}
        \item Node relabel $a \rightarrow b$: The node with label $a$ is changed to $b$.
        \item Node deletion $a \rightarrow \varnothing$: The node $\vv$ with label $a$ is removed, 
        and all children of $\vv$ become the children of the parent of $\vv$.
        \item Node insertion $\varnothing \rightarrow b$: The node $\vv$ with label $b$ is inserted,
        and some children of another node $\uu$ are made to be the children of $\vv$. 
\end{itemize}
We can define a cost function 
$c: (\Sigma \cup \{\varnothing \}) \times (\Sigma \cup \{\varnothing \}) \rightarrow \RR^{+} \cup \{0 \}$ 
that assigns a non-negative value to every edit operation.    
For a sequence $S = (s_i)_{i=1}^m$, where $s_i$ is an edit operation, the cost function 
is extended to this operation as $c(S) = \sum_{i=1}^m c(S)$. 
The edit distance between two trees $\TT_1$ and $\TT_2$ is defined as the cost 
of the minimal sequence of the edit operations from $\TT_1$ to $\TT_2$, that is
\begin{align}
        d_e(\TT_1, \TT_2) = \underset{S}{\text{min}}\ c(S).
\end{align}
An interesting property of the edit distance is that if the cost function $c$
satisfies the metric properties (definiteness, positivity, symmetry, and triangle inequality)
on $\Sigma \cup \{\varnothing \}$
then $d_e$ is also a metric on the space of rooted ordered labeled tree. 

The edit distance between two trees is an optimization 
over the space of all possible edit sequences, which is impossible to calculate. 
Indeed, an edit sequence transforming a tree $\TT_1$ to $\TT_2$ gives rise to 
a one-to-one, ancestor-preserving mapping between the nodes of the two trees.
This mapping is called \emph{edit mapping}.
Changing the order of the edit operations in the sequence does not change the mapping.
It is shown that the cost of the mapping coincides with the cost of the edit sequence. 
Thus, algorithms for computing the edit distance typically search for the mapping with 
the lowest cost instead.
\section{Tracking Methods for Merge Trees}
\label{sec-TrackingMethods}

In this section,
we present
the methods
that we selected for comparison.
Our focus lies
on methods
that track merge trees
\changed{using edit distance}
and therefore
provide a mapping
between the respective critical points
of two adjacent time steps.
\changed{These methods have been popular in recent years.
Hence, a dedicated comparison is timely.}





Before
going into the details,
we 
\changed{note that there are many feature tracking methods}
using topological descriptors 
\cite{yan2021scalar}.
\changed{These include}
critical points \cite{theisel03b, weinkauf10a, tricoche02, garth04b},
vortex core lines \cite{bauer02, theisel05b, weinkauf07c, Guenther15SciVis},
persistent diagrams \cite{soleretal2018, soler2019ranking}, 
extremum graphs \cite{narayanan2015distance, das2024time},
(nested) sublevel sets \cite{silverwang1997, silverwang1998, lukasczyketal2017, koeppweinkauf2019, lukasczyketal2020},
Reeb graphs \cite{edelsbrunner2004time, weber11, chen2013topology}, 
contour trees \cite{sohn06, lohfinketal2020, lohfinketal2021},
and, of course, 
merge trees, which we will focus on in the following.

\paragraph{Constrained Edit Distance $d_e$ by Sridharamurthy et al. \cite{sridharamurthy2018edit}}

Computing the edit distance between 
two unordered, labeled trees is NP-complete \cite{zhang1992}.
To enable computational feasibility,
Zhang \cite{zhang1996constrained} introduced the \emph{constrained edit distance},
which requires that disjoint subtrees are mapped to disjoint subtrees.
Sridharamurthy et al. \cite{sridharamurthy2018edit}
proposed to use the constrained edit distance for merge trees,
since merge trees can be viewed as ordered labeled trees,
where labels may reflect properties such as persistence.

Consider two merge trees $\TT_1 =(V_1, E_1)$ and $\TT_2 = (V_2, E_2)$ and two nodes
$\vv_1 \in V_1$, $\vv_2 \in V_2$.
Sridharamurthy et al. \cite{sridharamurthy2018edit}
defined the cost of the edit distance between 
the nodes as the $L_\infty$ between the birth and death of $\vv_1$ and $\vv_2$. 
The authors also introduced an overhang cost model for the edit operations. 
However, their experiments are exclusively performed with the $L_\infty$ cost model. 

\paragraph{Wasserstein Distance $d_w$ by Pont et al. \cite{pont22}}
The $L_2$-Wasserstein distance between two merges trees $\TT_1$ and $\TT_2$
is considered by Pont et al. \cite{pont22}.
The edit cost model is similar to the one by Sridharamurthy et al. \cite{sridharamurthy2018edit}.
However,
the method considers the cost of editing the nodes between two branch decomposition trees,
replacing $L_\infty$ with $L_2$.
They also added one more constraint for the edit mapping
such that if a root of a subtree is mapped to the empty symbol, 
so is the subtree.

\paragraph{Branch Mapping Distance $d_b$ by Wetzels et al. \cite{wetzels2022branch}}
While previous methods focus 
on the node-level mappings of merge and branch decomposition trees,
this approach considers the branches for the edit mapping. 
The mapping is defined such that it preserves the parent-child and ancestor-descendant
relationships between mapped branches. 
The branch mapping distance is defined analogously to the classic edit distance 
as the cost of the minimal branch mapping between two trees. 
The cost for the mapping can be defined as the persistence of the branch.
This method does not depend on a fixed branch decomposition.
It has a higher computational cost comparing to the other edit distance methods.

\paragraph{Path Mapping Distance $d_p$ by Wetzels and Garth \cite{wetzels2022path}}
The previous method
has been extended
to take into account generic paths in a tree.
They defined a different set of edit operations that is based on 
the deformation characteristic of merge trees, namely shrinking and extending an arc.
For instance, instead of simply deleting a node when removing an edge,
only the parent node is kept; and if that node has only one child left,
it will also be removed.
The new edit distance, called \emph{deformation-based edit distance},
is defined with the path edit sequences similarly to the normal edit distance.
This new distance is shown to be NP-hard \cite{wetzels2023stable}.
It is 
more tractable to constrain the edit operations to the leaves of the trees,
which is called the $1$-degree deformation-based edit distance or \emph{path mapping distance}.
Although having different theoretical properties,
the path mapping and branch mapping distances demonstrate algorithmic similarities.
More recently, Wetzels et al. \cite{wetzels2025stable} introduced a look-ahead parameter
to further stabilize the path mapping distance.

\paragraph{Other Related Methods}
Further methods for tracking merge trees exist,
but have not been selected for this comparison.
We focused on methods with publicly available implementations
to avoid a potential misinterpretation by us,
and we can only include a limited number of methods to begin with.
Nonetheless,
we want to mention some other interesting methods
for tracking merge trees and related tasks.
Oesterling et al. \cite{oesterling15} introduced time-varying merge trees.
Their method identifies all possible structural changes
between two consecutive time steps.
Beketayev et al. \cite{beketayev2014} proposed a distance measure that minimizes the cost of 
mapping branches between merge trees.
Saikia and Weinkauf \cite{saikia16a} constructed a directed acyclic graph
out of the merge trees of all time steps
and introduced global considerations into the feature tracking problem.
Köpp and Weinkauf \cite{koepp22}
lay out the temporal development of merge trees in a static layout
with minimal reliance on actual tracking information.
Li et al. \cite{li2025ot} used the theory of optimal transport 
to produce a partial matching between the leaves of two merge trees,
which allows for probabilistic tracking graphs.
Other approaches to comparing merge trees
include interleaving distances \cite{yan2022geometry, gasparovic2025intrinsic},
graph neural networks \cite{qin2025fast},
and local sensitive hashing \cite{lyu2025lsh}.

\paragraph{Ground Truths for Feature Tracking}
\changed{The creation of ground truths for feature tracking
is a challenging task.
In general, two principal approaches 
can be used: benchmark data generation 
and ground truth feature tracking results.
In this paragraph, we restrict our attention 
to one representative work from each
of the two categories.
Nilsson et al. \cite{nilsson2022towards} 
introduced a framework 
for generating predictable time-dependent scalar fields.
These scalar fields are accompanied 
by ground-truth
feature evolutions 
that enable systematic comparison
with the output of tracking methods. 
However, since the framework focuses on 
general feature tracking 
of extrema, it does not directly cater to 
merge tree-based feature tracking.   
In the other direction, 
analyzing the structural changes 
of a topological descriptor between 
consecutive time steps can also be 
regarded as a form of ground truth feature tracking.
Edelsbrunner et al. \cite{edelsbrunner2004time}
classified the combinatorial changes 
of Reeb graphs over time. 
By employing Jacobi curves,
they were able to update the Reeb graph
to reflect its temporal changes without recomputing 
it entirely from scratch for each time step.
This work has the potential to be a
standard for feature tracking.
However, adaptions are required to make it 
suitable for merge tree-based feature tracking.
We therefore leave this extension to future work.
}

\section{Experiments}
\label{sec-experiments}
In this section, we compare the tracking results  
produced by the four edit distance-based methods 
named in Section \ref{sec-TrackingMethods} empirically.

\subsection{Data Sets}
\label{sec-Datasets}
We introduce in this section the chosen data sets 
to compare the \changed{selected} merge tree-based tracking methods.
Due to 
\changed{page limitations}, 
we include here only the data sets
whose results exhibit the most qualitative differences.
More results can be found in the supplementary material.

The first data set is a time-dependent 2D flow field around a cylinder, 
referred to as \dataset{Cylinder2D}. 
It has been simulated numerically using Gerris Flow Solver \cite{germer11a}
by Weinkauf and Theisel \cite{weinkauf10c}. 
This data set is a vector field describing a 2D viscous flow around a cylinder
captured over 1001 time steps.
The flow forms the von \changed{K\'arm\'an} vortex street,
which is a repeating pattern of swirling vortices. 
We derive a scalar field from the velocity magnitude of each time step of this vector field
and use the join trees to track the center of the vortices.

\begin{figure}[t]%
\begin{subfigure}[t]{0.4\linewidth}%
\centering
\includegraphics[width=0.8\linewidth]{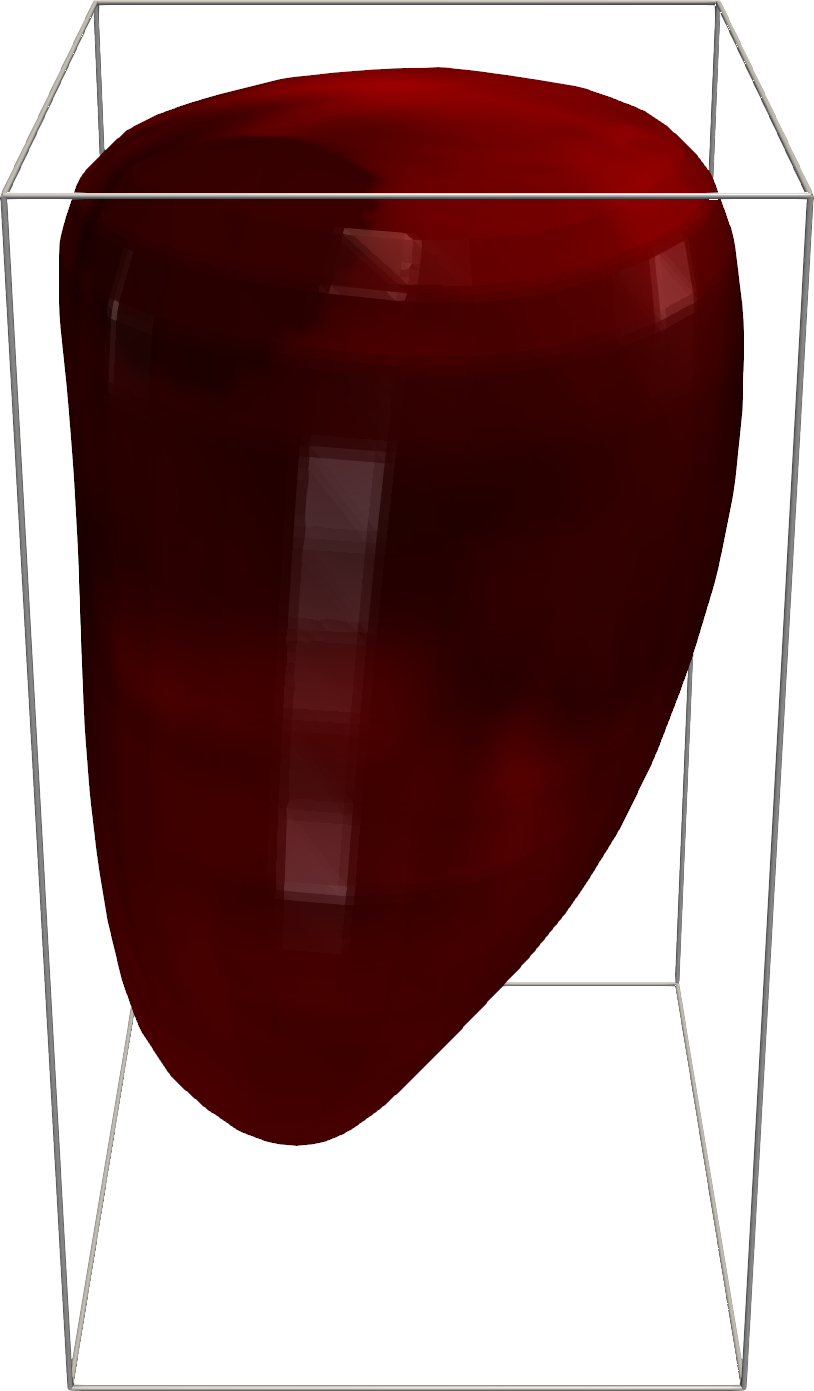}%
\caption{Time step 20.}%
\label{fig-HeartBeat3DT4336}%
\end{subfigure}
\begin{subfigure}[t]{0.4\linewidth}%
\centering
\includegraphics[width=0.8\linewidth]{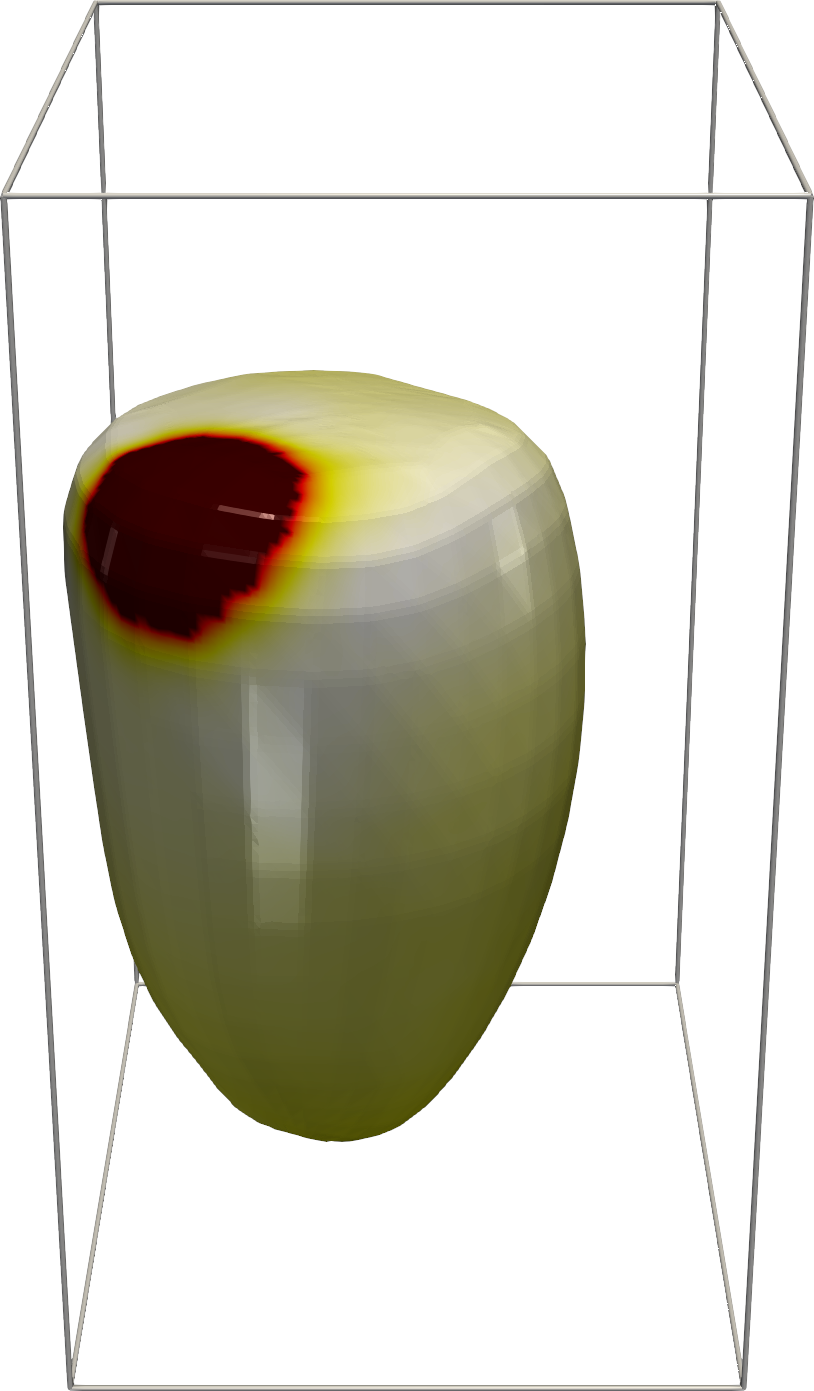}%
\caption{Time step 300.}%
\label{fig-HeartBeat3DT4616}%
\end{subfigure}%
\begin{subfigure}[t]{0.2\linewidth}%
\includegraphics[width=0.5\linewidth]{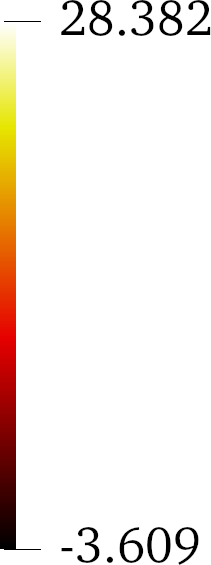}%
\end{subfigure}%

\caption{Pressure field of two time steps of \dataset{HeartBeat3D}.}%
\label{fig-HeartBeat3D}%
\end{figure}

Additionally,
we use a data set coming from a different application,
called \dataset{HeartBeat3D}. 
This data set models the left ventricular blood flow 
over different cardiac cycles 
using the FEniCS-HeartSolver \cite{larsson2017heart}.
It contains multiple properties for the simulation such as 
velocity, pressure, and magnitude of the triple decomposition. 
A sequence of 400 consecutive time steps of the pressure field of this data set 
is selected for our experiments. 
\changed{For our analysis, we focus on the pressure field,
where each time step is a 3D scalar field.}
Two different time steps of this data set are shown in \Cref{fig-HeartBeat3D}.
We use the split trees to track the regions with elevated pressure,
which can be used as an indicator of the flow patterns of the blood in the data set.

As these data sets contain noise, which can hinder the performance
of the tracking algorithms, we employ small simplification thresholds 
on these data set. The thresholds are 
1\% for \dataset{Cylinder2D},
and $0.5\%$ for \dataset{HeartBeat3D}. 
All experiments are conducted 
using TTK \cite{tiernyetal2018} on a machine using an Intel Core i9 chip with 64 GB of RAM.
We thank Wetzels et al. \cite{wetzels2025stable} for making their implementation 
of branch mapping and path mapping distances publicly available.

\subsection{Stability of the Tracking Algorithms}
\label{sec-stability}
\changed{These data sets provide the basis for our evaluations.
Certain properties of the data can 
have significant effects on the results.
In the following sections, we discuss the factors
that alter the tracking outcomes of the four algorithms
and assess their stability.
}

\subsubsection{Vertical and Horizontal Instability}
\label{sec-verticalHorizontal}
Saikia et al. \cite{saikia14a} characterized two types of instabilities that 
can affect the matching between two merge trees.

\paragraph{Vertical instability} Suppose that we have a merge tree with two major branches 
corresponding to the two largest maxima with very similar values. 
One branch has more sub-branches, indicating richer features,
while the other is relatively simpler.
\changed{Now, consider a very similar merge tree to this one}.  
Intuitively, we would prefer to map the nodes
between the branches with similar structure complexity, that is 
feature-rich branches with one another and likewise for simpler branches.
However, a small perturbation in the function value can reverse
the order of the maxima, making the feature-less branch more dominant. 
Thus, mappings that rely on branch persistence or value differences can 
fail to match the nodes as desired.
This problem is particularly pronounced in branch decomposition-dependent methods.  

\paragraph{Horizontal instability} This type of instability is often referred 
to as 
\changed{\emph{saddle swap}}. 
Consider a merge tree with two saddle points that
are located very close together. Under a small functional perturbation, their positions 
can be swapped. Consequently, this difference in the positions of the saddle points can alter 
the matching between two merge trees. This issue poses a challenge for most 
methods that are based on edit distance, 
as there are often constraints on the edit operations.
\Cref{fig-instabilityTypes} demonstrates two types of instability.

\definecolor{minColor}{HTML}{8080E6}
\definecolor{saddleColor}{HTML}{FFF380}
\definecolor{maxColor}{HTML}{E68080}

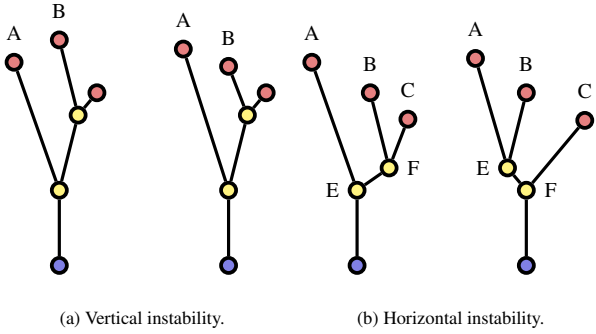
\begin{figure}
\begin{subfigure}{\ltwopicwidth}
\centering
\begin{tikzpicture}[
    level distance=1.0cm,
    sibling distance=0.5cm,
    every node/.style={
        circle, draw, minimum size=2mm, inner sep=0pt, 
        font=\footnotesize,
        fill=white, 
        line width=0.5mm,
        label distance=2mm,
    },
    mininumNode node/.style args={#1}{
        fill=minColor,
        label={[label distance=0mm]below:#1}
    },
    saddleNode node/.style args={#1}{
        fill=saddleColor,
        label={[label distance=0mm]right:#1}
    },
    maximumNode node/.style args={#1}{
        fill=maxColor,
        label={[label distance=1mm]above:#1}
    },
    edge from parent/.style={
        draw,
        line width=1.2pt
    },
    grow=up
]

\node[mininumNode node={}] (root1){}
    child { node[saddleNode node={}] (uSaddle1) {}
        child { node[saddleNode node={}] (bSaddle1) {}
            child { node [maximumNode node={}, yshift=-20pt] (smallMax1) {}}
            child { node [maximumNode node={B}] (mainMax1_1) {}}
        }
        child { node[maximumNode node={A}, yshift=20pt, xshift=-10pt] (mainMax2_1) {}}
    };

\node[mininumNode node={}, right = 2cm of root1] (root2){}
    child { node[saddleNode node={}] (uSaddle2) {}
        child { node[saddleNode node={}] (bSaddle2) {}
            child { node [maximumNode node={}, yshift=-20pt] (smallMax2) {}}
            child { node [maximumNode node={B}, yshift=-10pt] (mainMax1_2) {}}
        }
        child { node[maximumNode node={A}, yshift=25pt, xshift=-10pt] (mainMax2_2) {}}
    };
\end{tikzpicture}
\caption{Vertical instability.}
\label{fig-veriticalInstability}
\end{subfigure}
\begin{subfigure}{\ltwopicwidth}
\centering
\begin{tikzpicture}[
    level distance=1.0cm,
    sibling distance=0.5cm,
    every node/.style={
        circle, draw, minimum size=2mm, inner sep=0pt, 
        font=\footnotesize,
        fill=white, 
        line width=0.5mm,
        label distance=2mm,
    },
    mininumNode node/.style args={#1}{
        fill=minColor,
        label={[label distance=0mm]below:#1}
    },
    saddleRightNode node/.style args={#1}{
        fill=saddleColor,
        label={[label distance=0.5mm]right:#1}
    },
    saddleLeftNode node/.style args={#1}{
        fill=saddleColor,
        label={[label distance=0.5mm]left:#1}
    },
    maximumNode node/.style args={#1}{
        fill=maxColor,
        label={[label distance=1mm]above:#1}
    },
    edge from parent/.style={
        draw,
        line width=1.2pt
    },
    grow=up
]

\node[mininumNode node] (root1){}
    child { node[saddleLeftNode node={E}] (uSaddle1) {}
        child { node[saddleRightNode node={F}, yshift=-20pt, xshift=5pt] (bSaddle1) {}
            child { node [maximumNode node={C}, yshift=-10pt] (smallMax1) {}}
            child { node [maximumNode node={B}] (mainMax1_1) {}}
        }
        child { node[maximumNode node={A}, yshift=20pt, xshift=-10pt] (mainMax2_1) {}}
    };

\node[mininumNode node, right = 2cm of root1] (root2){}
    child { node[saddleRightNode node={F}] (uSaddle2) {}
        child { node[maximumNode node={C}, yshift=-2pt, xshift=15pt] (mainMax2_2) {}}
        child { node[saddleLeftNode node={E}, yshift=-20pt] (bSaddle2) {}
            child { node [maximumNode node={B}, yshift=0pt] (smallMax2) {}}
            child { node [maximumNode node={A}, yshift=13pt, xshift=-5pt] (mainMax1_2) {}}
        }
    };
\end{tikzpicture}
\caption{Horizontal instability.}
\label{fig-horizontalInstability}
\end{subfigure}
\caption{
    Shown are two types of instability.
    In the left figure, nodes $A$ and $B$ 
    change their values, 
    causing the other to become 
    the new global maximum, 
    which modifies the BDT. 
    Methods that rely on BDTs
    struggle with this type of instability.
    The right figure illustrates the horizontal instability.
    Saddle nodes $E$ and $F$ have very close values. 
    A small perturbation can swap their positions 
    and consequently change the structure of the tree.
    This type of instability affect most methods.
}
\label{fig-instabilityTypes}
\end{figure}

One way to mitigate this horizontal instability is to employ a
step called $\epsilon$-processing \cite{thomas11}.
This approach merges two saddle points in a bottom-up manner such that 
if the difference in functional values of those points 
are less than a threshold $\epsilon$, they will be merged.
The parameter $\epsilon$ is often scaled to $[0,1]$ according to the data range.
At $\epsilon = 1$, all extrema are connected to a single merged saddle point.
As the $\epsilon$-process further collapses the tree after the persistent simplification,
the resulting tree may not reflect the original definition of merge trees.
Furthermore, this process does not resolve the instability problems, it merely conceals them.
Thus, in all subsequent experiments, we exclude this process.

Wetzels et al. \cite{wetzels2023stable} showed that
using an unconstrained edit distance 
can improve the robustness of the matching against both types of instabilities.
However, as with the case of the classic unconstrained edit distance, this problem 
is NP-complete, making it impractical to compute.
To address this challenge, Wetzels et al. \cite{wetzels2025stable} recently proposed a 
heuristic for their previously introduced path mapping distance.
This new method can effectively handle 
horizontal instability up to a user-specific look-ahead parameter.

\begin{figure*}

\begin{center}
\includegraphics[width=0.3\linewidth]{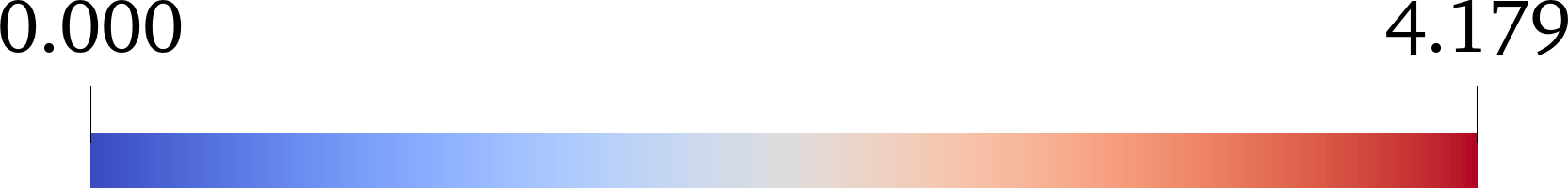}
\end{center}
\hfill

\begin{subfigure}[b]{\twopicwidth}
\includegraphics[width=\linewidth]{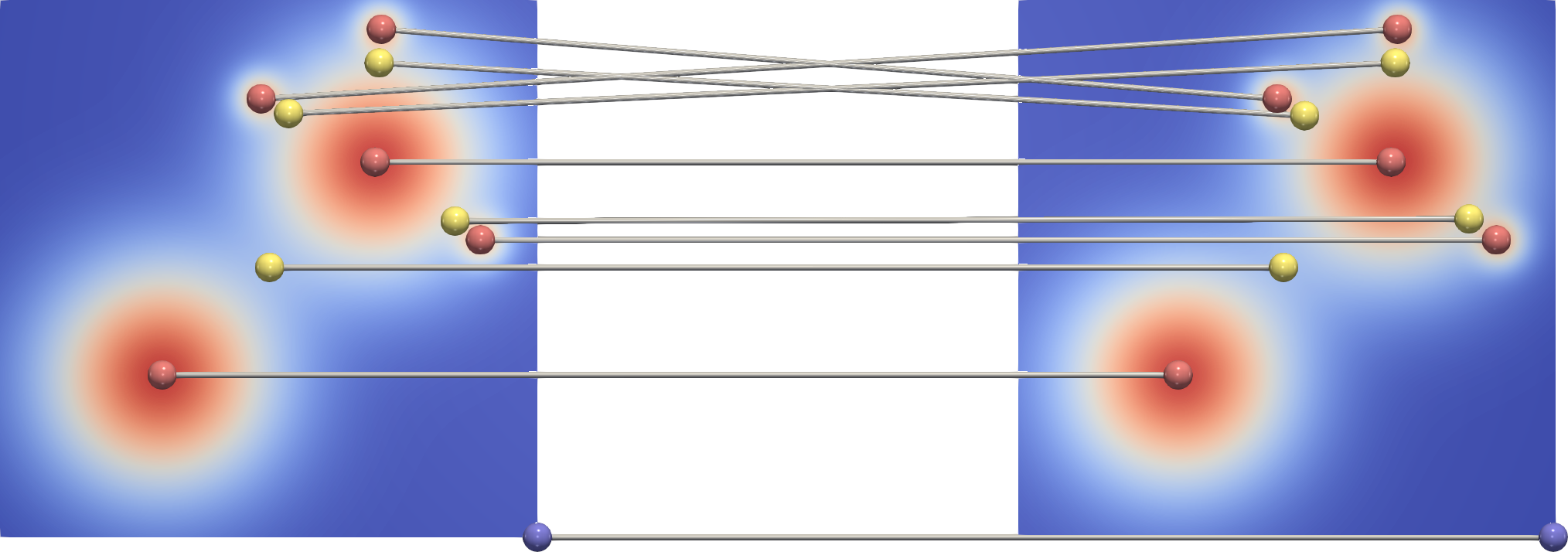}
\caption{Wasserstein distance \cite{pont22}.}
\label{fig-instablityWasserstein}
\end{subfigure}
\hfill
\begin{subfigure}[b]{\twopicwidth}
\includegraphics[width=\linewidth]{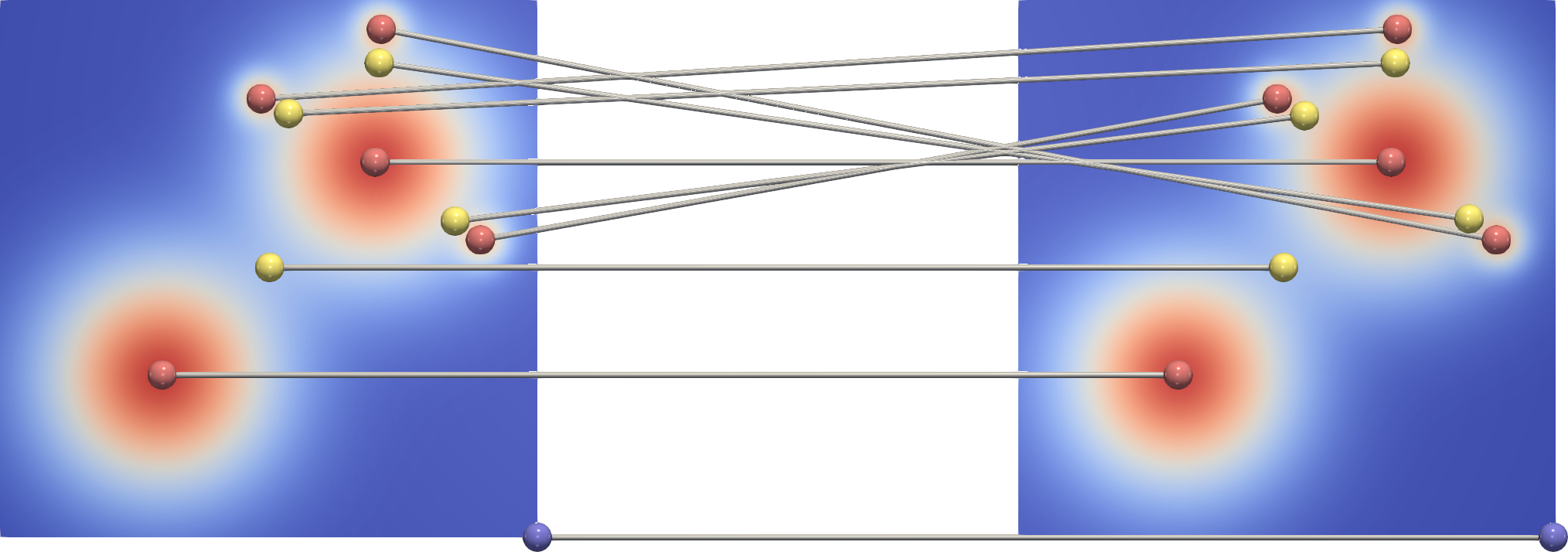}
\caption{Constrained edit distance \cite{sridharamurthy2018edit}.}
\label{fig-instablityEdit}
\end{subfigure}
\hfill
\begin{subfigure}[b]{\twopicwidth}
\includegraphics[width=\linewidth]{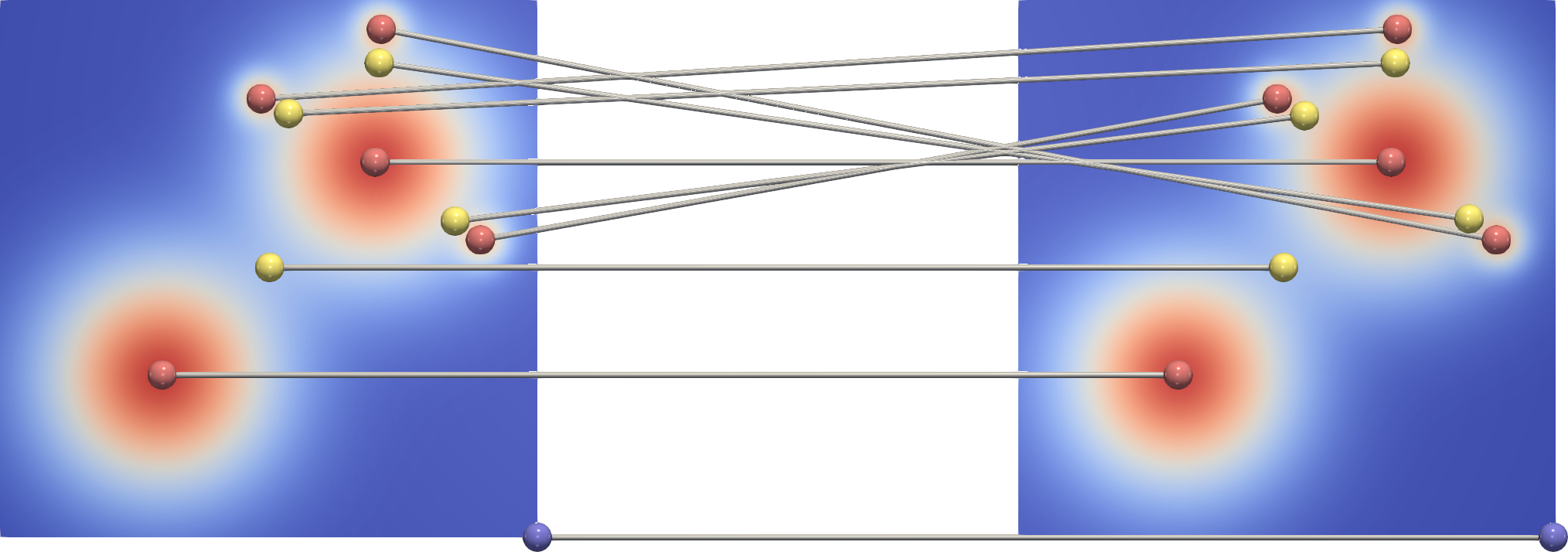}
\caption{Branch mapping distance \cite{wetzels2022branch} and path mapping distance \cite{wetzels2022path}.}
\label{fig-instablityBranch}
\end{subfigure}
\hfill
\begin{subfigure}[b]{\twopicwidth}
\includegraphics[width=\linewidth]{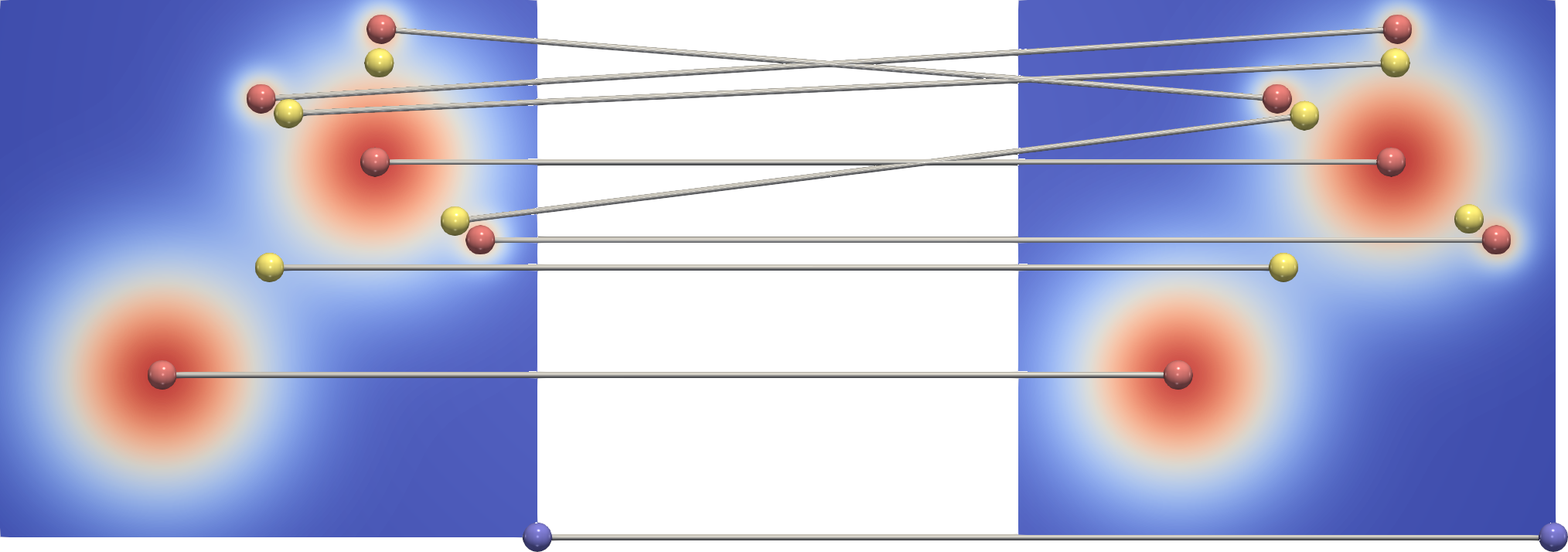}
\caption{Path mapping distance \cite{wetzels2025stable} with look-ahead of $10$, equal to the depth of the trees.}
\label{fig-instablityPath4}
\end{subfigure}
\caption{Similar features determined by the chosen methods between 
two very similar scalar fields.
The fields differ slightly, which trigger the instabilities between two merge trees.
There is no difference between the results of the constrained edit distance \cite{sridharamurthy2018edit},
branch mapping distance \cite{wetzels2022branch} and path mapping distance \cite{wetzels2022path}.
All chosen methods, even for the path mapping distance \cite{wetzels2025stable} with a high look-ahead,
are affected by these instabilities and 
fail to produce the expected result, 
that is all horizontal lines should be parallel.
}
\label{fig-instability}
\end{figure*}

\paragraph{Experiments} 
We highlight the impact of the instabilities with different tracking algorithms. 
To do so, we construct a base scalar field consisting of three 
Gaussian peaks with similar size and height, 
each positioned at a different location. 
From one of these peaks, we generate two branching features of the same size. 
A similar procedure is applied to another peak, 
but to create three small features. 
We randomly perturb the base scalar field by slightly 
varying the positions, heights, and size of the larger peaks 
as well as the small features within the range $[-0.01, 0.01]$.
This process simultaneously triggers both types of instabilities.
We expect that the methods can compute the correspondences between
the peaks and smaller features.

The matching results between the chosen methods are shown in \Cref{fig-instability}
of two randomly chosen fields. We can see that there are differences between the matching 
between the methods and neither of them produces the expected result.
This behavior can be observed not only for these two instances but other similar scalar fields as well.
Another observation that can be seen is that with a look-ahead of $10$, 
the path mapping distance did not produce a matching for a pair of saddle points. 
Indeed, as we later observed, the saddle points are often less stable than the local extrema, 
and methods often sacrifice the saddle-saddle connections 
for higher-quality correspondences between the extrema.
This issue can be problematic when tracking saddle features is also of interest.   

\subsubsection{Stability with respect to Noise}
\label{sec-stableNoise}
Having seen the main reasons for the divergence in the tracking paths, 
we now examine their stability in a controlled noisy data set. 

To assess stability, we are interested in how long each method
can track the original features as the noise level increases. 
Our goal is to create a noisy data set in which the original features
can be verified within the noisy data set. 
Therefore, instead of randomly perturbing the function values,
our strategy to create noisy data sets is described in the next paragraphs.

\begin{figure}
\begin{subfigure}[b]{0.8\linewidth}
\begin{subfigure}[b]{\linewidth}
\includegraphics[width=\linewidth]{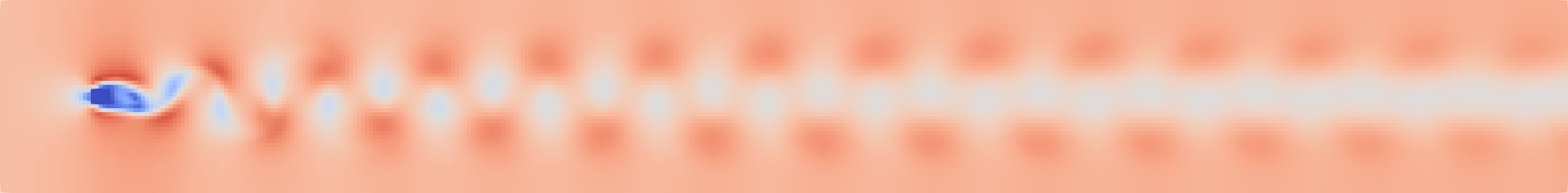}
\caption{Original scalar field.}
\label{fig-StableNoiseOrg}
\end{subfigure}
\begin{subfigure}[b]{\linewidth}
\includegraphics[width=\linewidth]{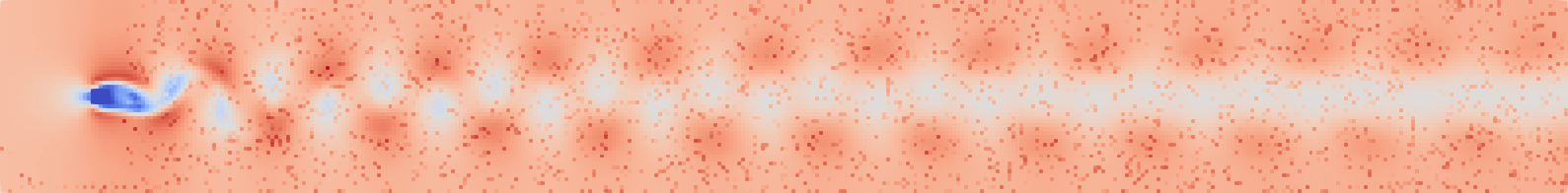}
\caption{Noisy scalar field with $\tau= 15\%$ and $\epsilon = 20\%$.}
\label{fig-StableNoiseT15E20}
\end{subfigure}
\end{subfigure}
\begin{subfigure}[b]{0.18\linewidth}
\centering
\includegraphics[height=2.9cm]{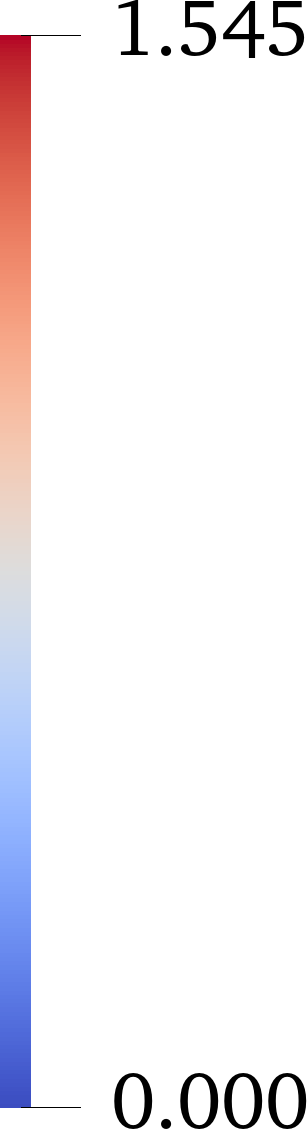}
\end{subfigure}
\caption{Noisy scalar field created from a cleaned scalar field using our strategy.
The noises are added such that the local minima in the merge tree of the noisy field 
includes the local minima of the cleaned field.  
}
\label{fig-StableNoise}
\end{figure}

Consider a scalar field $f$ with the corresponding merge tree $\TT = (V,E)$,
for any $\vv \in V$, let $\idx{\vv}$ be the index of the vertex of $\vv$ in $f$.
We want to add noise to $f$ to create $f'$ with the merge tree $\TT' = (V', E')$
such that the extrema of $f$ is a subset of $f'$, 
i.e., $\idx{\leaves{\TT}} \subseteq \idx{\leaves{\TT'}}$.
We can achieve this by taking advantage of the Morse cells \cite{forman98}.
Around $\tau \%$ points are chosen randomly from the vertices set of $f$.
For each chosen point $\pp$, it will be modified such that the value 
of each point will not be larger than the maximum (for the case of ascending manifold)
or smaller than the minimum (for the case of descending manifold) of the Morse cell
$\pp$ belongs to. The level of the noise will be monitored by a parameter $\epsilon$, 
represent the percentage of the data range. 
An example of the noisy scalar field produced from the introduced approach
using a time step from \dataset{Cylinder2D}
can be seen in \Cref{fig-StableNoise}.

For a fixed $\tau$, as $\epsilon$ increases, we will obtain a series of data sets
in which noise gradually becomes more significant. 
For $t$ increasing values of $\epsilon$, this series can be interpreted as a time-dependent data set, 
from which a family of merge trees $(\TT_i)_{i=1}^t$ can be computed. 
Consider a tracking algorithm $D$, 
a tracking path $p = (\vv_1,\dots,\vv_k)$ that track an original node of $\TT_1$ 
is a path such that $\vv_i \in \TT_i$ and $\idx{\vv_i} = \idx{\vv_j}$ for all $1\leq i \neq j \leq t$.
In other words, $p$ starts from the first tree and all nodes in $p$ has the same index. 
The tracking stability of the extrema of $D$ is
\begin{align}
    s(D)_\ell = \dfrac{\sum_{\vv \in \leaves{\TT_1}} \len{\vv}}{t|\leaves{\TT}|},
\end{align}    
where $\len{\vv}$ is the length of the tracking path starting from $\vv$.
The denominator in the stability formula represents the hypothetical most stable tracking method
where it can track the original extrema despite the added noise. 
The more stable a method is, the longer it can follow the original nodes
from the original tree, and the closer the stability is to $1$. 

\begin{figure}[t]
\centering
\includegraphics[width=0.8\linewidth]{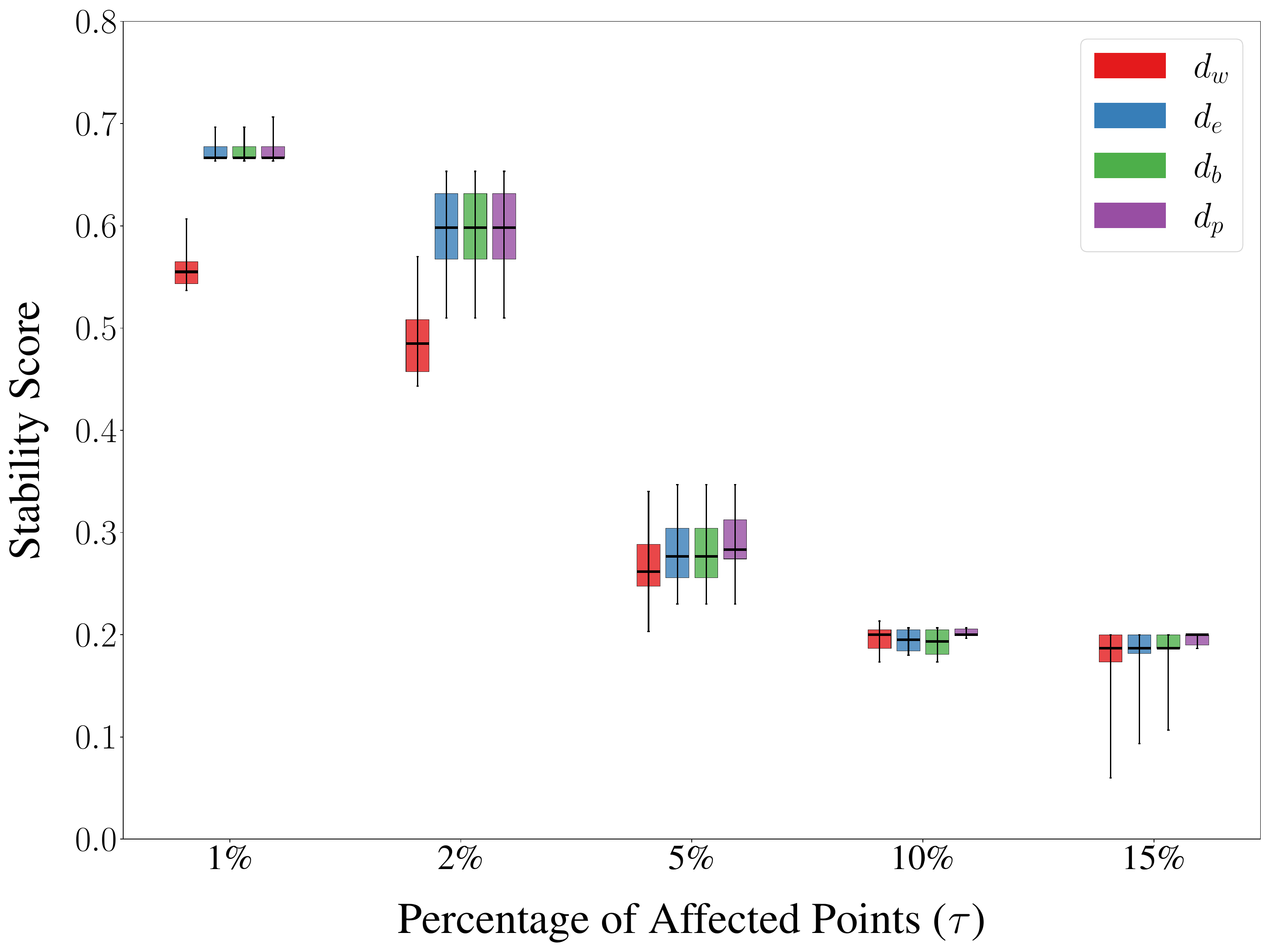}
\caption{
    The tracking stability of the extrema of four methods 
    If we stop tracking at $\epsilon_\text{max} = 5\%$.
    The median, first and third quantiles, maximum and minimum 
    values of the stability scores at each level 
    of affected points are shown for each method. 
    The stability score decreases as the percentage 
    of affected points and the level of noise increase. 
    In general, $d_w$ produces the tracking 
    with the lowest scores, while the other three methods behave quite similarly. 
    The path mapping distance $d_p$ with look-ahead of $4$ 
    yields slightly better and more stable scores compared to $d_e$ and $d_p$. 
}
\label{fig-StableNoiseResults}
\end{figure}

\paragraph{Experiment} 
We use the introduced score to determine 
the stability of the tracking results produced by 
the four algorithms.
We choose time step 500 from \dataset{Cylinder2D}.
For each $\tau \in \{1\%, 2\%, 5\%, 10\%, 15\% \}$, 
we generate $40$ noisy scalar fields with
increasing noise level $\epsilon$ uniformly sampled between $0\%$ and $\epsilon_\text{max}\%$. 
Due to the randomness of the added noise, 
we repeat the same process $10$ times for each $\tau$. 
We set the look-ahead parameter of $d_p$ to $4$ for this experiment as well 
as subsequent experiments, 
which provide a reasonable stability
and computational time as reported by its authors \cite{wetzels2025stable}.
The results for $\epsilon_\text{max}$ are reported in \Cref{fig-StableNoiseResults}. 

%
Overall, these methods exhibit strong stability in the presence of small amount of noise.
For instance, when roughly $1\%$ of the points are perturbed
with noise at $5\%$ of the data range,
stability scores can reach up to $0.7$.
From our experiments, 
even under more substantial noise levels, up to $\epsilon_\text{max} = 20\%$, 
most methods still maintain scores around $0.5$.
The decline in stability is gradual as the noise level
and the impact becomes more significant.
We can see that as the percentage of affected noise increases,
the stability scores of all four methods 
converge to a small value. 
The tracking results produced by $d_w$ show 
the most distinct stability scores, often lower than the other three methods. 
This behavior may stem from the fact that $d_w$ works directly
on BDTs, and it imposes additional constraints on the edit mapping
between them. 
Particularly,
the constraint is such that a 
subtree will collapse if its root is mapped to an empty node. 
Thus, in noisy environments, the likelihood of nodes 
not find their correspondences increases, 
leading to reduced stability scores.
Interestingly, $d_e$ and $d_b$ have scores that are very similar in our experiments,
despite $d_e$ is not a deformation-based method. 
%
Additionally, the implementations of $d_p$ and $d_b$ are similar
despite differences in theoretical motivations, 
which result in comparable stability scores in most cases.
The branch decomposition $d_p$ shows slightly higher stability,
thanks to the stability control parameter, look-ahead.

\begin{figure*}[t]
\begin{subfigure}[b]{\twopicwidth}
    \centering
    \includegraphics[width=\linewidth]{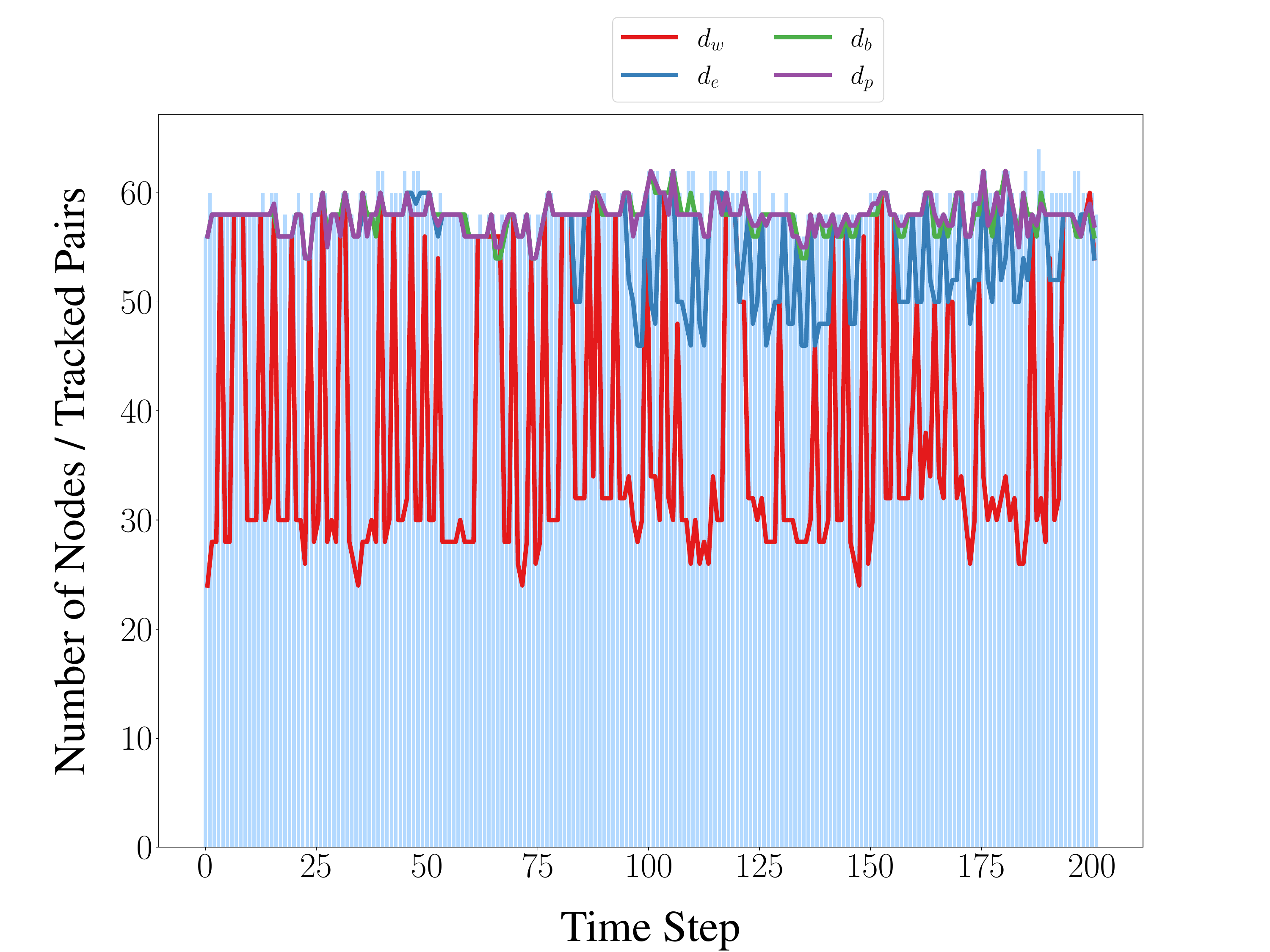}
    \caption{First $200$ time steps of \dataset{Cylinder2D}.}
    \label{fig-cylinder2DDist}
\end{subfigure}
\begin{subfigure}[b]{\twopicwidth}
    \centering
    \includegraphics[width=\linewidth]{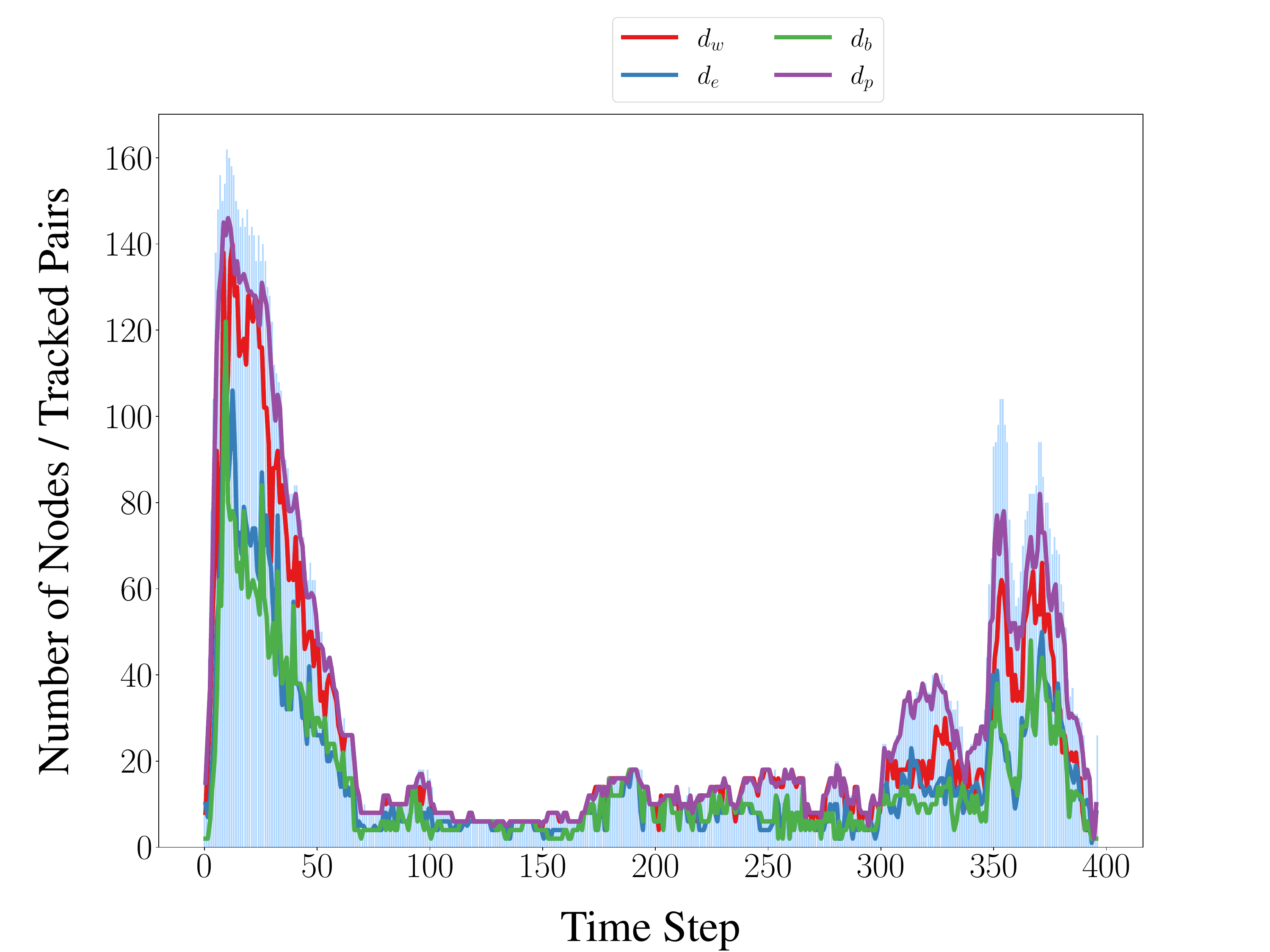}
    \caption{\dataset{HeartBeat3D}.}
    \label{fig-heartBeat3DDist}
\end{subfigure}
\caption{Number of tracked nodes (colored and labeled) 
    relative to the total nodes of merge trees (light blue bars)
    from the chosen time steps of the data sets.
    The number of nodes of the merge trees show patterns of the data sets.
    }
\label{fig-MatchingPairDistribution}
\end{figure*}

\subsection{Tracking on Real-World Data Sets}
\label{sec-RealDatasets}
We are now turning our focus on the differences
using real-world data sets introduced in \Cref{sec-Datasets}.
We analyze the performance of the four methods in the context of feature tracking.
Individual performance of these methods on each data set is considered. 
Additionally, we calculate the pairwise differences between the tracking paths.

\subsubsection{Individual Performance}
\label{sec-individualPerformance}
For individual performance, we consider two factors. 
First, we examine the number of matched nodes between two consecutive time steps. 
This number reflects the local properties 
of the tracked paths and can reveal latent patterns of the data.
Second, we investigate the length of the tracked paths,
which served as a global property. 
This measure provides insights into the number of features in the data 
and highlights how effectively each method 
tracks persistent features over time. 
A meaningful tracking should be capable of capturing
both local and global properties to some extent.
The results are discussed in the following paragraphs.

%

\paragraph{Distributions of Matched Nodes}
We show the number of matched nodes between consecutive time steps
for the first $200$ instances of \dataset{Cylinder2D}
and for the chosen $400$ time steps of \dataset{HeartBeat3D}
in \Cref{fig-cylinder2DDist} and \Cref{fig-heartBeat3DDist}, respectively.
%
The merge trees for \dataset{Cylinder2D}
have the numbers of nodes vary periodically,  
representing new vortices being born
and old vortices moving out of the domain.   
This data set exhibits periodic properties: 
a half period spans approximately $38$ time steps
and a full period occurs roughly every $75$ time steps \cite{wetzels2025stable}.
For the case of \dataset{HeartBeat3D}, 
we can roughly divide
the data set into three phases: 
very strong activity during the first 60 time steps, 
a resting period over the next 250 time steps,
and increased activity again for the remaining time.  

In ideal scenarios, we would want to have the methods to 
find correspondences for almost every pair of nodes between two merge trees after simplifying.
It is not the case for any of the methods in either of the data sets. 
In both cases, $d_p$ determines the most tracked pairs between two trees
among the four methods. 
With \dataset{Cylinder2D}, $d_b$
follows $d_p$
while in \dataset{HeartBeat3D}, $d_w$
take over the role of the $d_b$
in the discrepancy with $d_p$.   

As established in \Cref{sec-stability}, $d_w$
is expected to produce the most different outcomes.
This expectation can be observed for \dataset{Cylinder2D},
where the numbers of tracked nodes produced by this method 
fluctuate tremendously. 
From this data set, initially, $d_e$
follows $d_b$ and $d_p$ closely
but begins to show greater variation around time step $78$,
though not as pronounced as $d_w$. 
Thus, the branch decomposition-independent methods 
can capture the periodic behavior of \dataset{Cylinder2D},
while the others cannot due to their instability.

The roles of the $d_w$ and the $d_b$
are swapped in \dataset{HeartBeat3D}.
The numbers of matched nodes produced by the methods
are relatively stable compared to their behaviors in \dataset{Cylinder2D},
which can partially be explained by the smoother structure of the pressure field 
in \dataset{HeartBeat3D}. 
All methods worked on this data set can capture three phases.
Apart from the discussed $d_w$ and $d_b$, 
we can see that the $d_e$
is very similar to the $d_b$.

\begin{figure*}
\begin{subfigure}[b]{\twopicwidth}
    \centering
    \includegraphics[width=\linewidth]{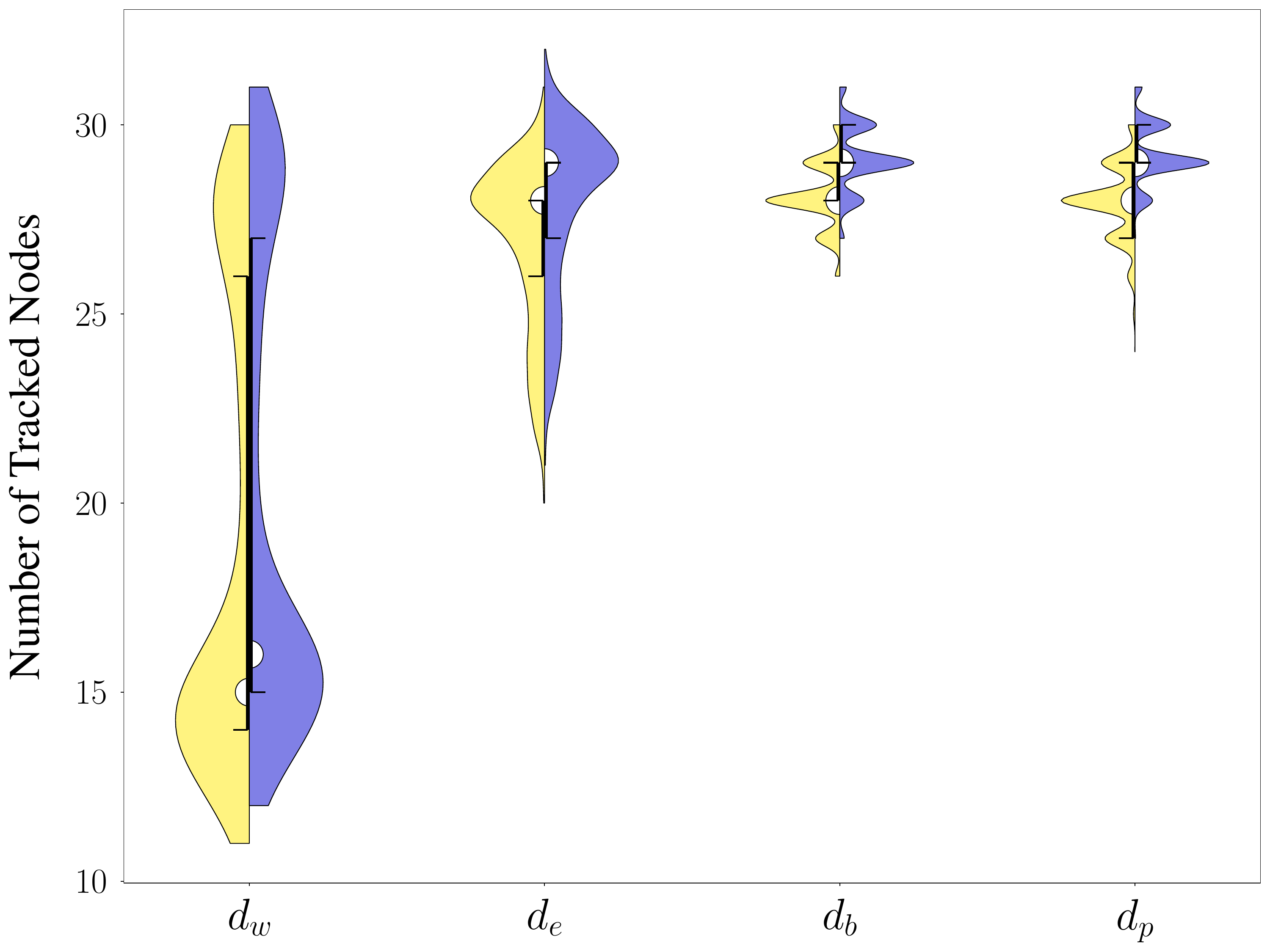}
    \caption{\dataset{Cylinder2D}.}
    \label{fig-cylinder2DViolin}
\end{subfigure}
%
%
\begin{subfigure}[b]{\twopicwidth}
    \centering
    \includegraphics[width=\linewidth]{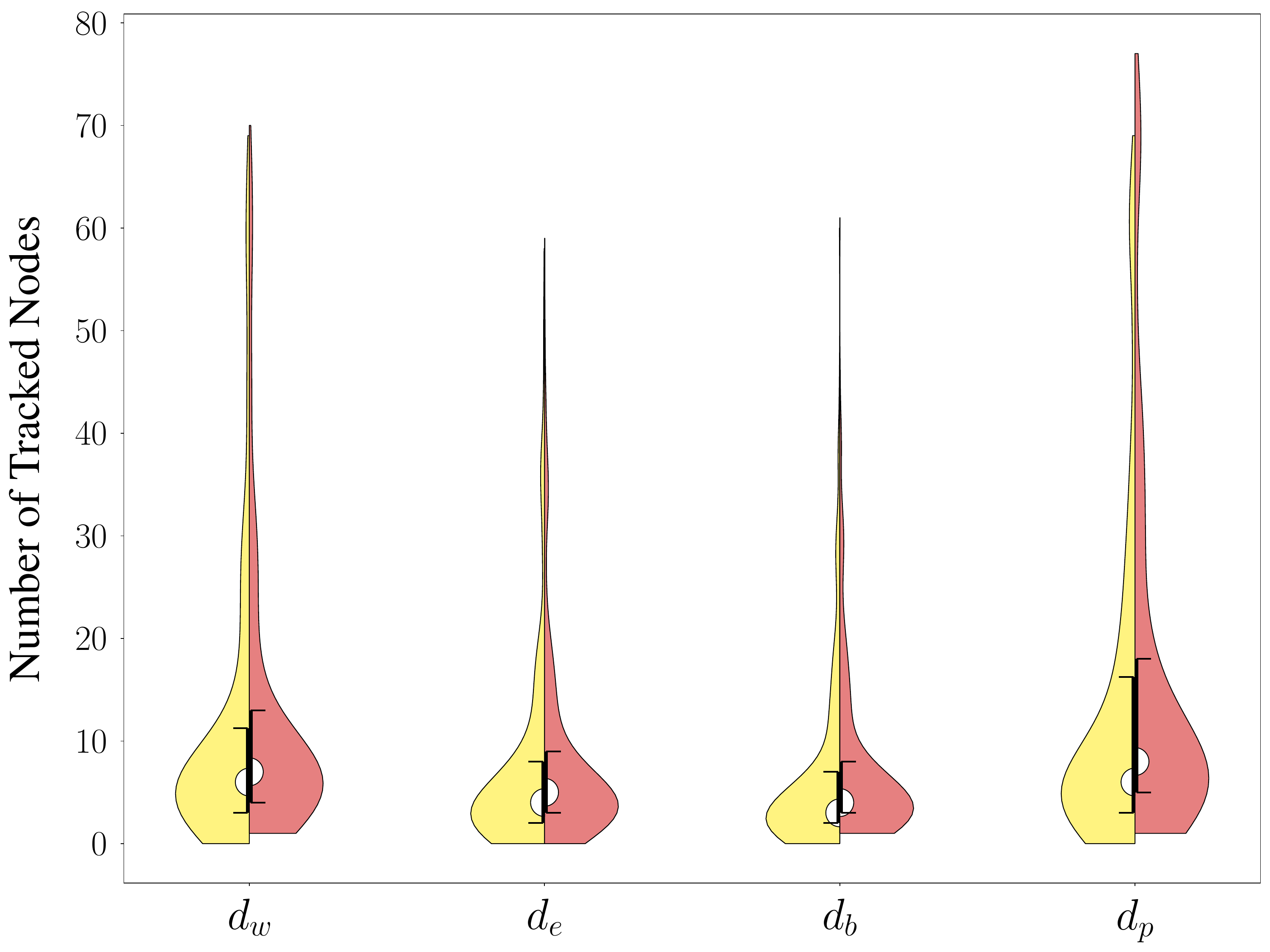}
    \caption{\dataset{HeartBeat3D}.}
    \label{fig-heartBeat3DViolin}
\end{subfigure}
\caption{Distributions of tracked nodes 
    over all time steps of two data sets of four methods,  
    divided by types:
    inner (yellow) and leaves (colored blue for minima
    and red for maxima).
    The first and third quantiles, together with the medians of 
    the distributions are also shown.
    Differences between the distributions can be inferred.
    }
\label{fig-violinCriticalPoints}
\end{figure*}

\paragraph{Characterized by Critical Types} 
When grouped by types of matched critical points,
the four methods display distinct distributions in \Cref{fig-cylinder2DDist},
whereas they have similar shapes in \Cref{fig-heartBeat3DViolin}.
There are more differences from $d_e$ to $d_w$,
partly because the latter method has a larger search space when computing
its version of edit distance.  
The matching distribution of the two branch decomposition-independent
methods are similar partly due to their similar implementations.
For \dataset{Cylinder2D}, the numbers of $d_w$
tends to be the lower-end, 
but it is similar to $d_p$
with \dataset{HeartBeat3D}: the body of the distribution is thinner. 
Furthermore, as mentioned in \Cref{sec-verticalHorizontal}, 
the more stable methods prioritize mappings between the extrema 
over the correspondences of saddle points,
which can also be seen in both graphs of \Cref{fig-violinCriticalPoints} as well.

\begin{figure}[t]%
\begin{subfigure}[t]{\ltwopicwidth}%
\includegraphics[width=\ltwopicwidth]{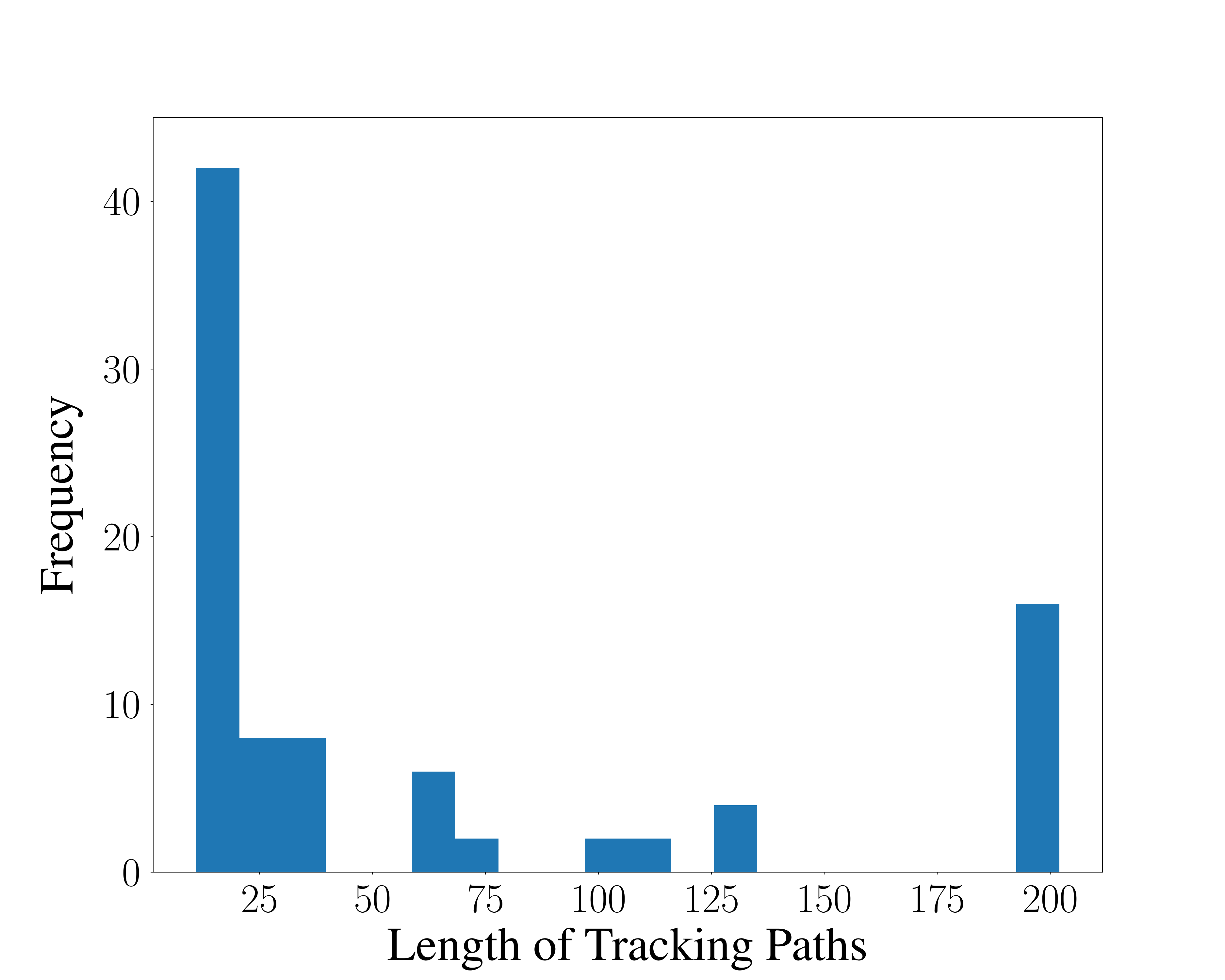}%
\caption{$d_w$.}%
\label{fig-LengthHistogramCylinder2DW}%
\end{subfigure}
\hfill%
\begin{subfigure}[t]{\ltwopicwidth}%
\includegraphics[width=\ltwopicwidth]{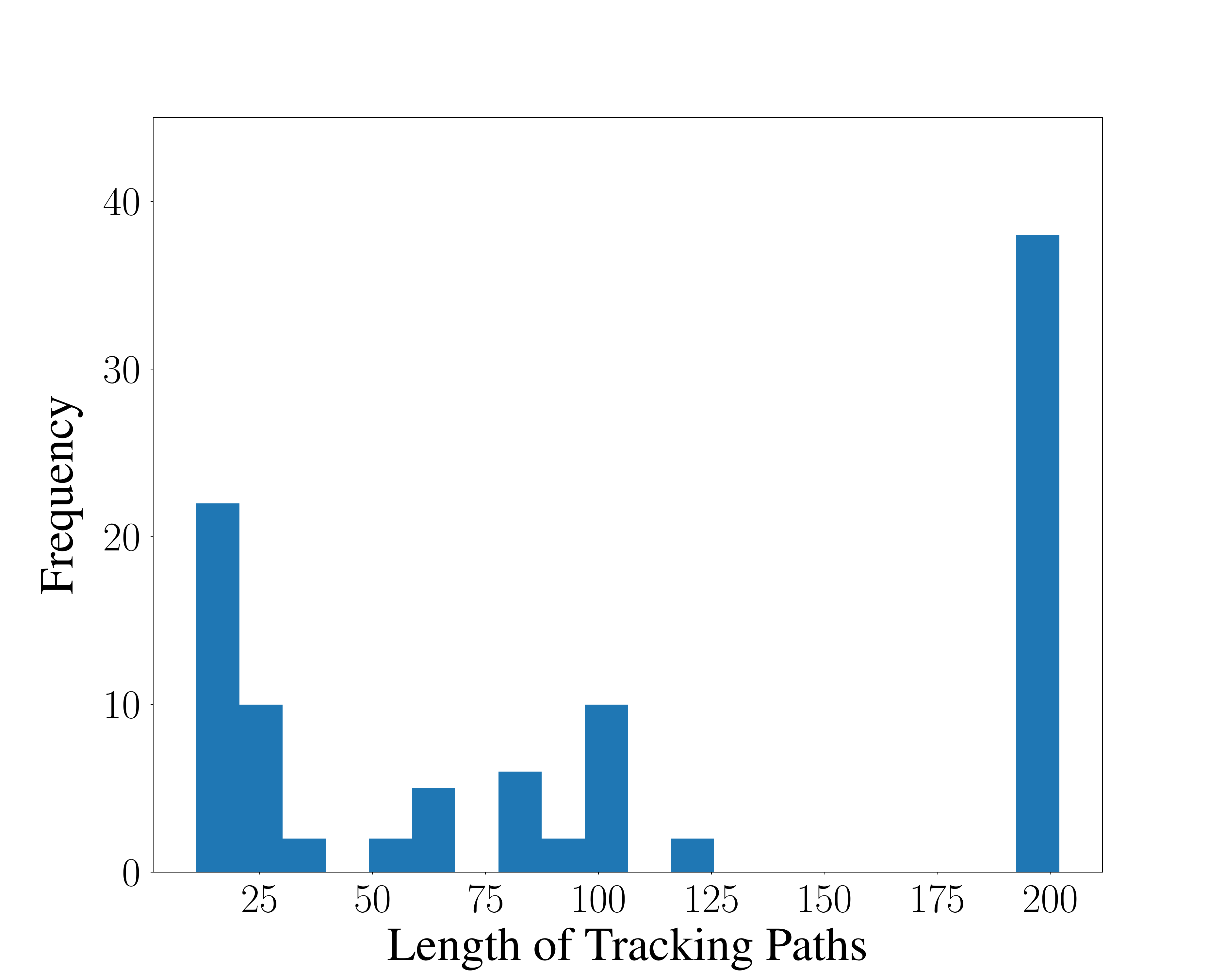}%
\caption{$d_e$.}%
\label{fig-LengthHistogramCylinder2DE}%
\end{subfigure}%
\\%
\begin{subfigure}[t]{\ltwopicwidth}%
\includegraphics[width=\ltwopicwidth]{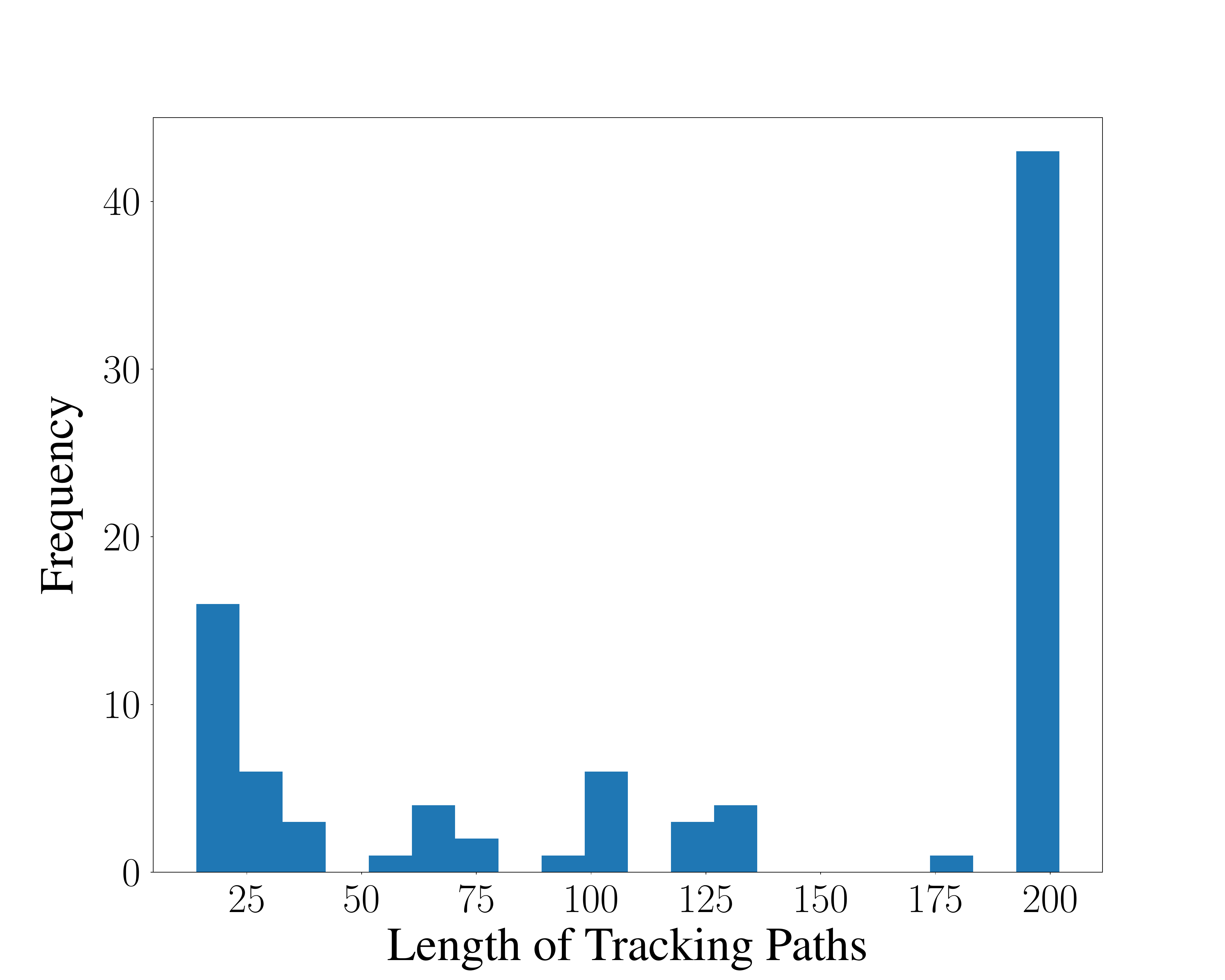}%
\caption{$d_b$.}%
\label{fig-LengthHistogramCylinder2DB}%
\end{subfigure}%
\hfill%
\begin{subfigure}[t]{\ltwopicwidth}%
\includegraphics[width=\ltwopicwidth]{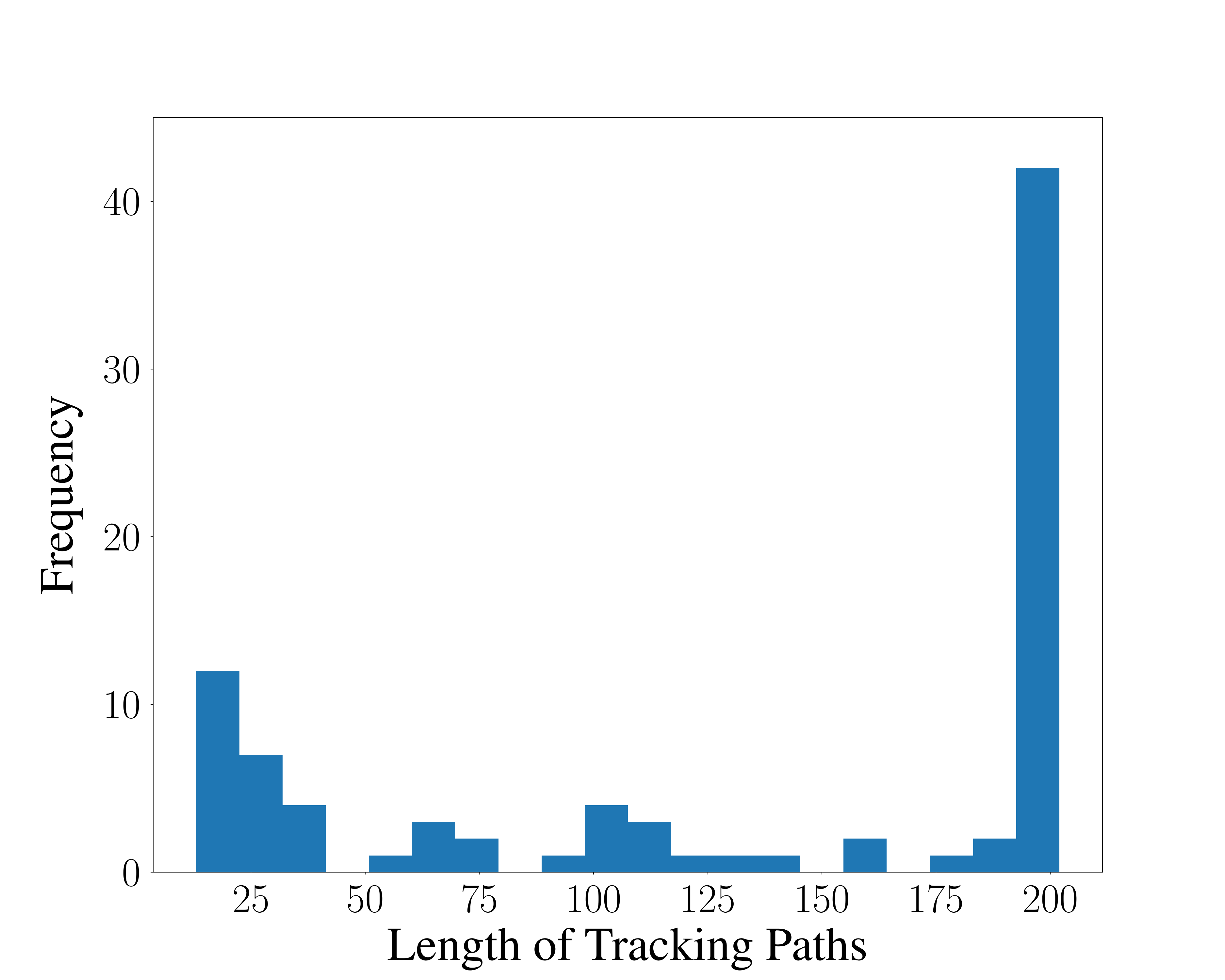}%
\caption{$d_p$.}%
\label{fig-LengthHistogramCylinder2DP}%
\end{subfigure}%
\caption{
    Histograms of the length of the tracking paths of the first $200$ instances 
    of \dataset{Cylinder2D}. 
    We removed the paths with length shorter than $10$ due to their dominance
    over other paths in the histograms. 
}%
\label{fig-LengthHisgoramCylinder2D}%
\end{figure}

\begin{figure}[t]%
\begin{subfigure}[t]{\ltwopicwidth}%
\includegraphics[width=\ltwopicwidth]{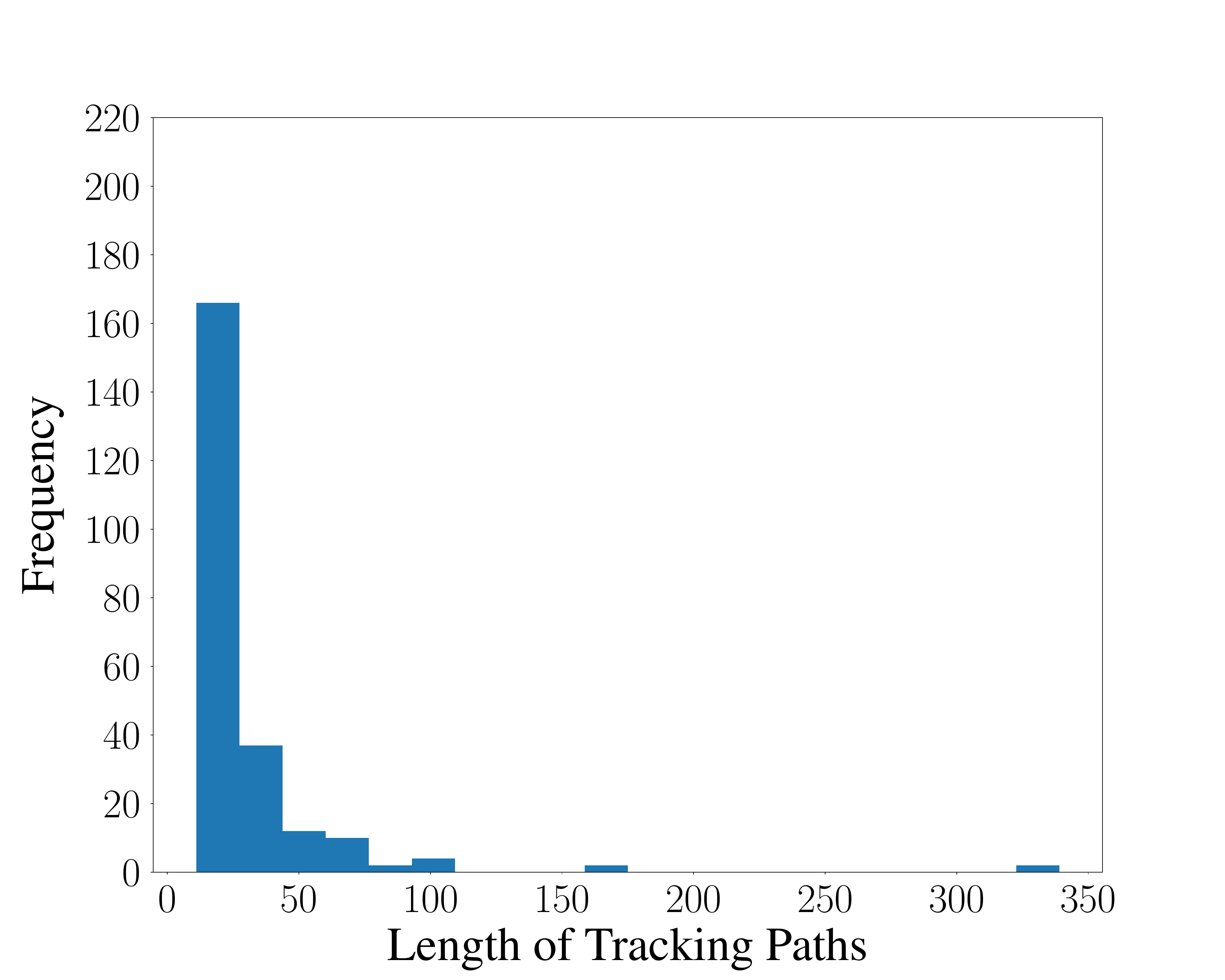}%
\caption{$d_w$.}%
\label{fig-LengthHistogramHeartBeat3DW}%
\end{subfigure}
\hfill%
\begin{subfigure}[t]{\ltwopicwidth}%
\includegraphics[width=\ltwopicwidth]{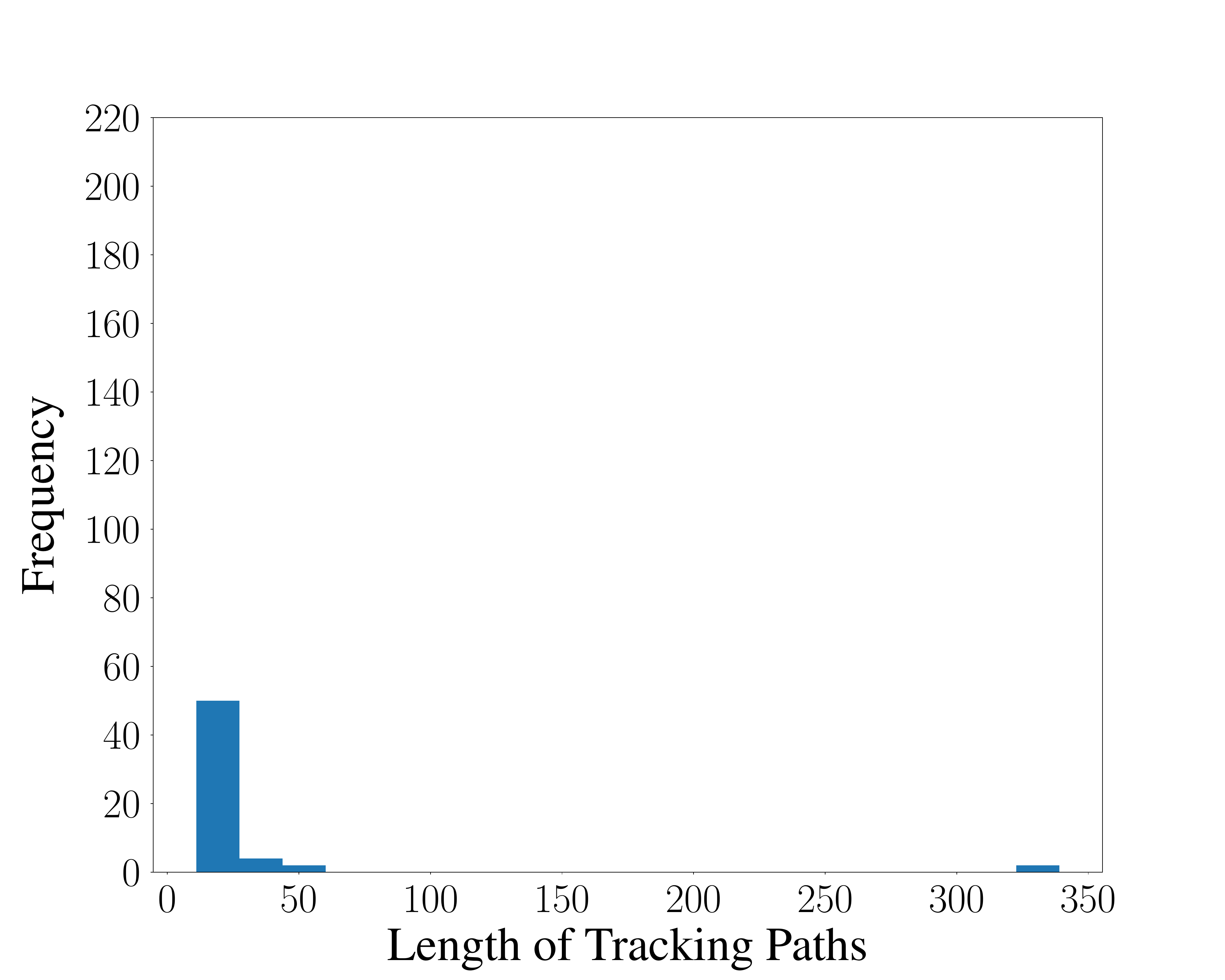}%
\caption{$d_e$.}%
\label{fig-LengthHistogramHeartBeat3DE}%
\end{subfigure}%
\\%
\begin{subfigure}[t]{\ltwopicwidth}%
\includegraphics[width=\ltwopicwidth]{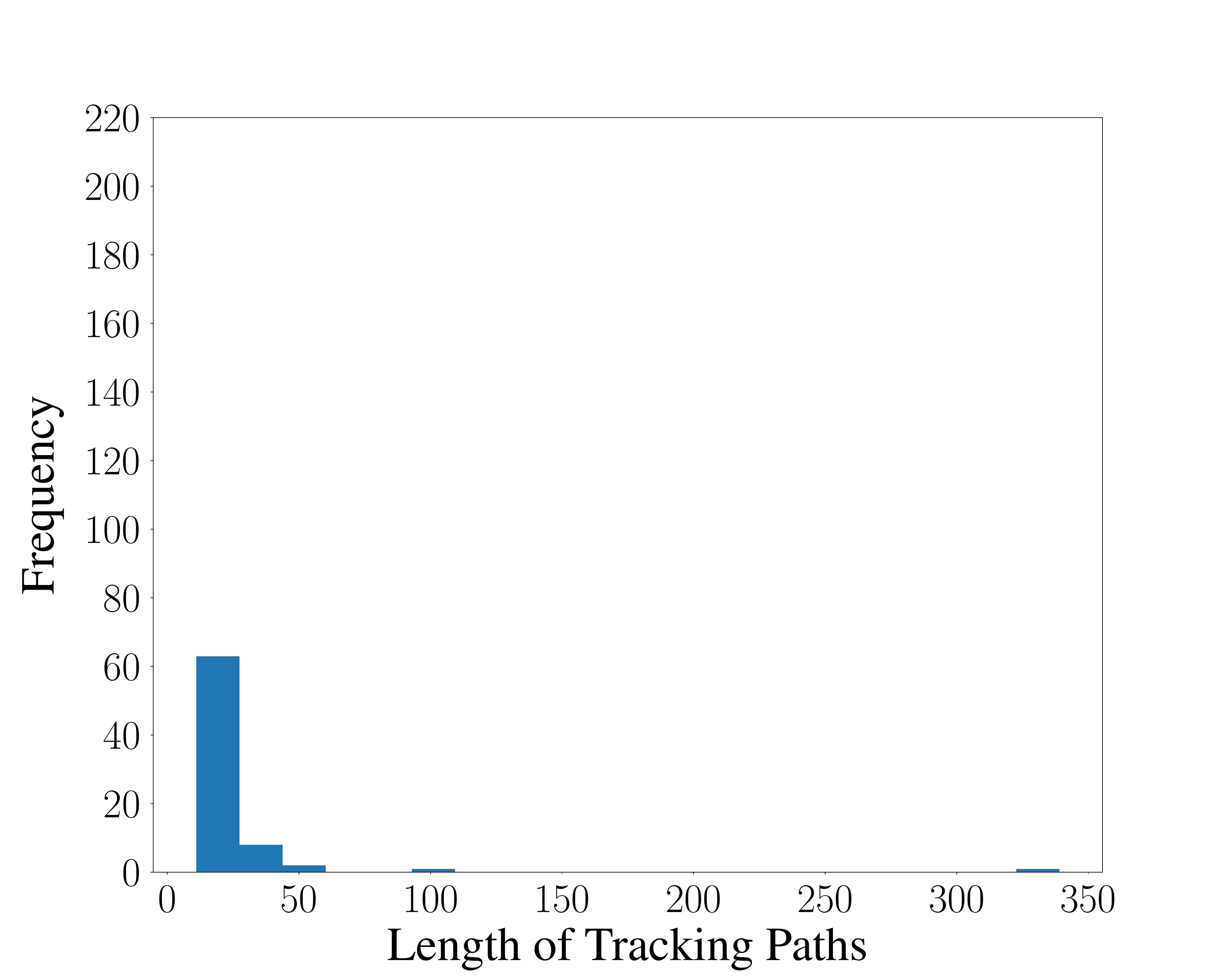}%
\caption{$d_b$.}%
\label{fig-LengthHistogramHeartBeat3DB}%
\end{subfigure}%
\hfill%
\begin{subfigure}[t]{\ltwopicwidth}%
\includegraphics[width=\ltwopicwidth]{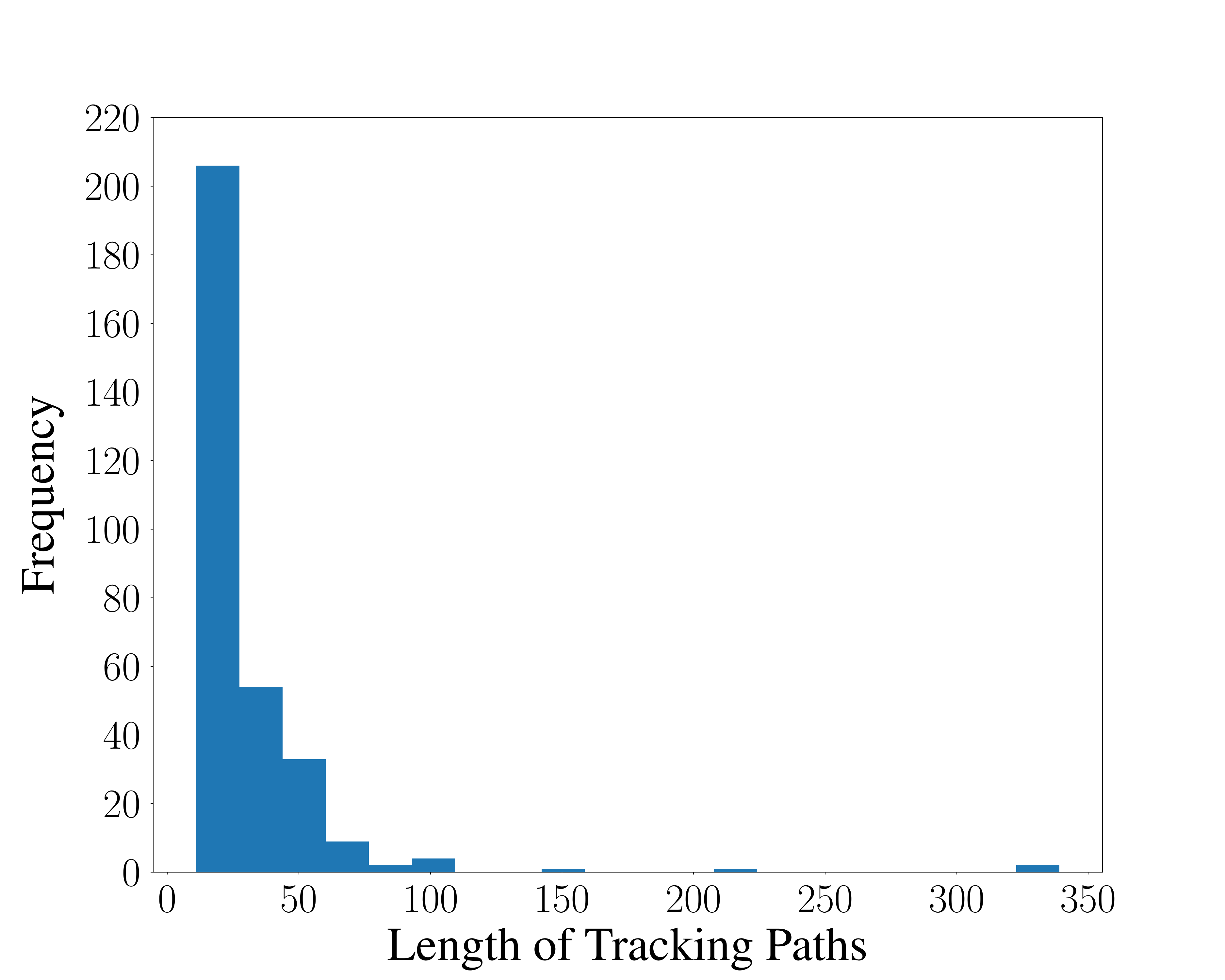}%
\caption{$d_p$.}%
\label{fig-LengthHistogramHeartBeat3DP}%
\end{subfigure}%
\caption{Histograms of the length of the tracking paths of the first $200$ instances 
    of \dataset{HeartBeat3D}.}%
\label{fig-LengthHisgoramHeartBeat3D}%
\end{figure}

\paragraph{Length of Tracking Paths}
\Cref{fig-LengthHisgoramCylinder2D} presents histograms of tracking path 
lengths derived from the initial $200$ time steps of \dataset{Cylinder2D}.
The four chosen methods detect enormous numbers of short-lived features.
Therefore, we removed the tracking paths whose length is shorter than $10$
(or $5\%$ of the total of time steps).
Even after discarding the tiny tracks, 
the histogram of $d_w$ still shows a strong presence 
of the shorter tracks, 
with less than $20$ tracks having the maximal length of $200$.
This trend reverses as we shift the focus to other methods.
We can see that as we move from $d_e$ 
to the two branch-independent methods, 
longer tracks increase and shorter tracks diminish.
This observation can indicate that the methods can 
determine more persistent features over time.  
On the other hand, \Cref{fig-LengthHisgoramHeartBeat3D} shows 
that all methods detect large numbers of short-lived features
with longer-lived ones being relatively scarce.
The reason for these computational patterns 
may lie in the nature of the data set.
As the blood traverses the heart, it can spread to the whole region,
as can be seen in \Cref{fig-HeartBeat3DT4336},
corresponding to a merge tree with numerous features.
Alternatively, the blood can concentrate on smaller areas
as illustrated in \Cref{fig-HeartBeat3DT4616}.
During transitions between these states, many nodes die,
explaining the overall shape of the histograms.   
Noticeably, the $d_p$
produces an abundance of short tracks, 
though we can not verify if all of them
are meaningful. 

\subsubsection{Pairwise Differences}
\label{sec-pairwiseDifferences}
\changed{Although a reliable ground truth is generally unavailable,}
we can still relate the results to each other.
\changed{A systematic comparison with non-merge tree-based
methods is left for further studies.
}
We rely on pairwise differences between the tracking results themselves.
Let $\TT_1 = (V_1, E_1)$ and $\TT_2 = (V_2, E_2)$ be two merge trees. 
The tracking result produce by a method $\DD$ is the set of tracked pairs 
$\DD(\TT_1, \TT_2) = \{ (\vv_1, \vv_2) | \vv_1 \in V_1, \vv_2 \in V_2\}$.
Thus, the pairwise difference between the tracking results
of method $\DD_1$ to method $\DD_2$ from $t_1$ to $t_2$ is
$|\DD_1(\TT_1, \TT_2) \backslash \DD_2(\TT_1, \TT_2)|$.
In other words, it is the number of tracked pairs that are produced by $\DD_1$ but not $\DD_2$.
The total difference of a data set with $k$ time steps 
can be calculated by taking the sum of the pairwise differences
over all possible consecutive time steps, that is
\begin{align}
    \pdiff{\DD_1, \DD_2} = \sum_{i=1}^{k-1} |\DD_1(\TT_{i}, \TT_{i+1}) \backslash \DD_2(\TT_{i}, \TT_{i+1})|.
\end{align}
Note that in general,
$\pdiff{\DD_1, \DD_2} \neq \pdiff{\DD_2, \DD_2}$.
An example of such cases is when a tracked pair is produced by $\DD_1$ but not by $\DD_2$.

\begin{figure}
\begin{subfigure}[b]{\ltwopicwidth}
    \centering
    \includegraphics[width=\linewidth]{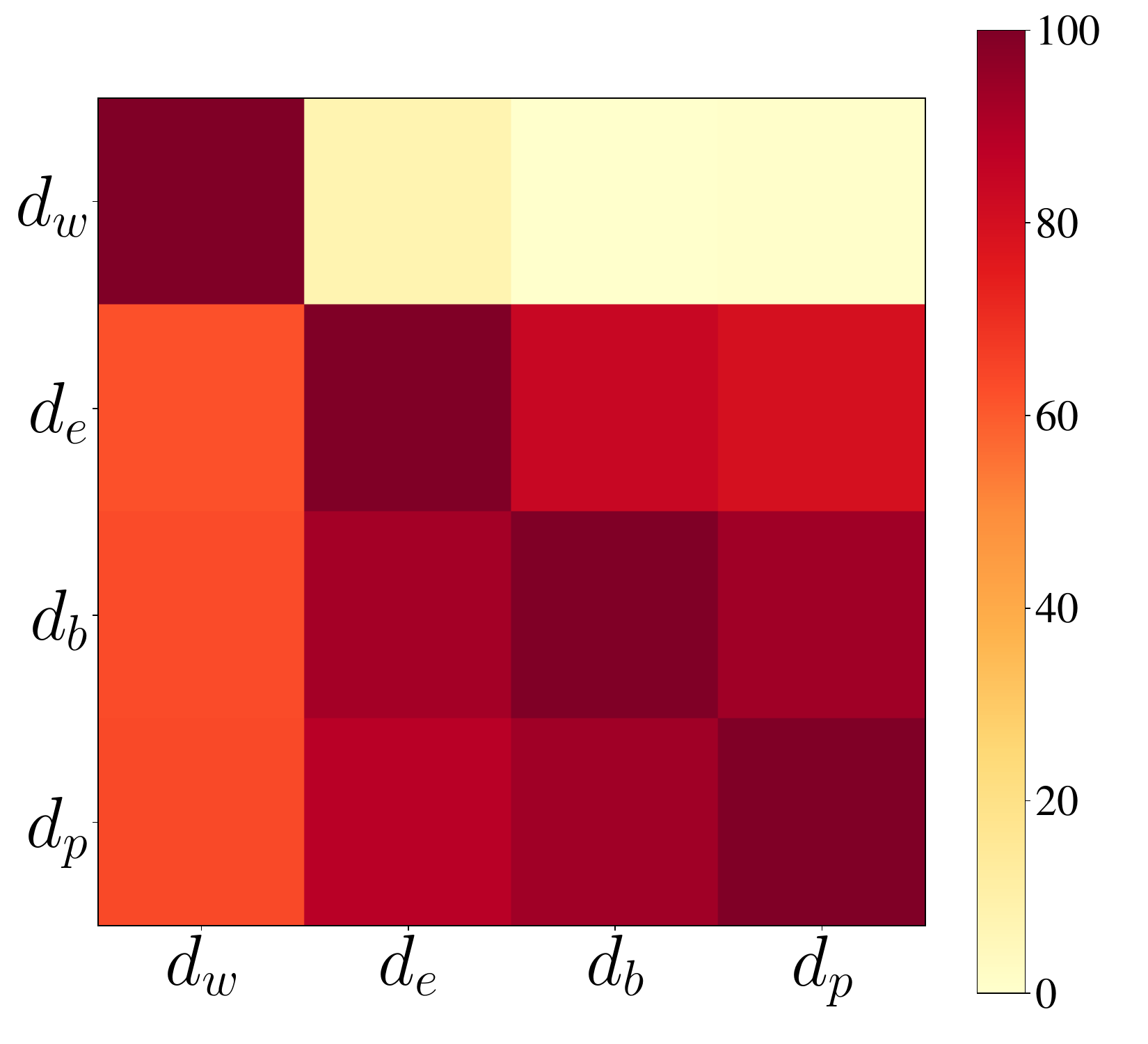}
    \caption{\dataset{Cylinder2D}.}
    \label{fig-cylinder2DSimilarity}
\end{subfigure}
\begin{subfigure}[b]{\ltwopicwidth}
    \centering
    \includegraphics[width=\linewidth]{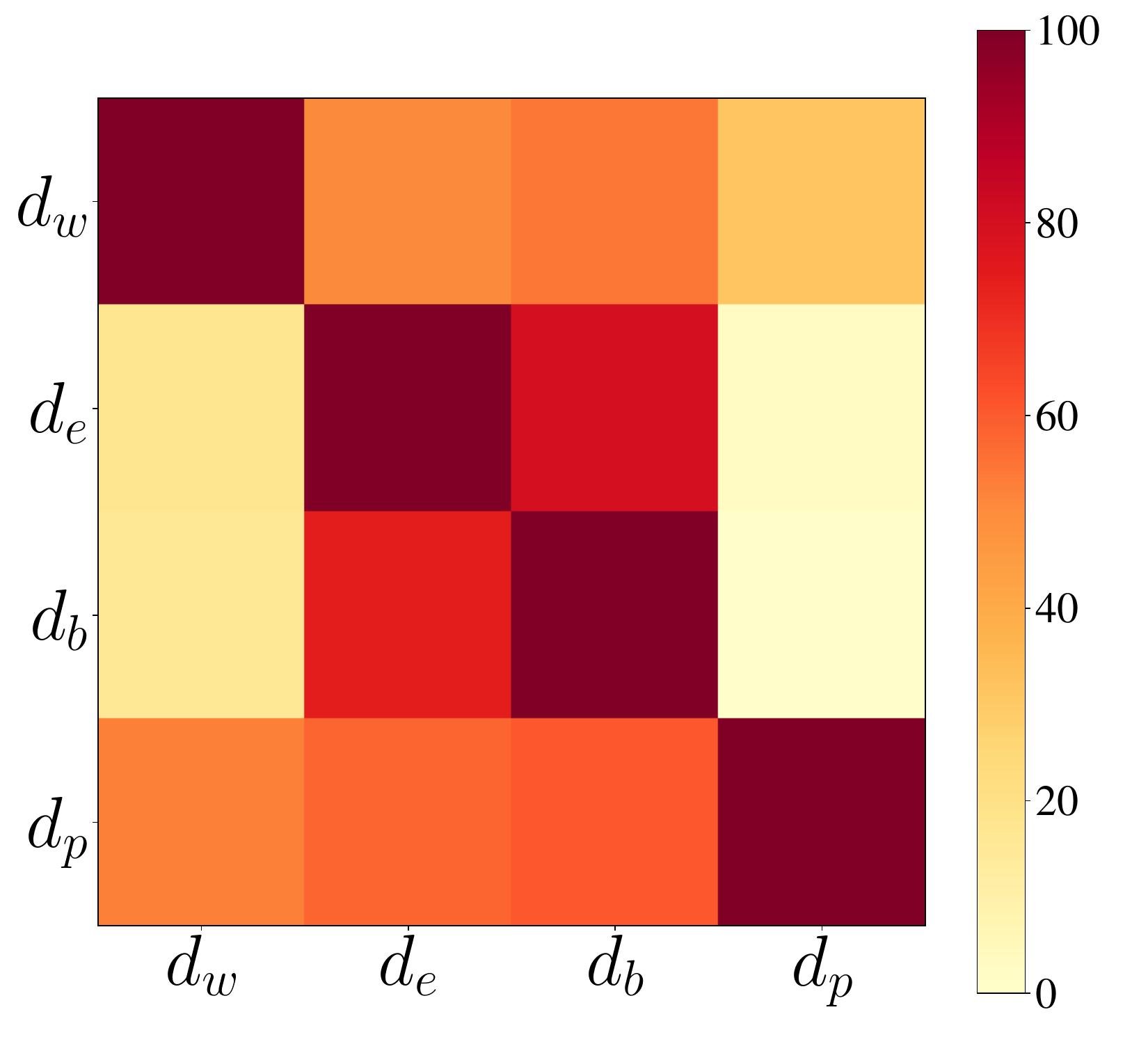}
    \caption{\dataset{HeartBeat3D}.}
    \label{fig-HeartBeat3DSimilarity}
\end{subfigure}
\caption{
    Pairwise differences between the methods on the columns
    in relation to the methods on the rows
    for the two data sets.
    The total numbers are scaled to be between $0\%$ and $100\%$
    to show their similarities.  
    }
\label{fig-SimilarityMatrices}
\end{figure}

\paragraph{Results}
The differences of the methods in the two data sets are further consolidated in 
the similarity matrices shown in \Cref{fig-SimilarityMatrices}.
We can see that methods other than $d_w$
form a cluster with \dataset{Cylinder2D}. 
The tracking result of $d_w$ in this data set
is largely dissimilar to the other three,
since it only shares around $20\%$ of the tracked pairs with the other.
The two methods $d_b$ and $d_p$
are closer to each other than either is to $d_e$. 
As for the remaining similarity matrix shown in \Cref{fig-HeartBeat3DSimilarity},
we can clearly see that there are two groups of tracking results: 
one group with $d_w$ and $d_p$,
the other group contains the remaining methods $d_e$ and $d_b$. 
The second group has more in common among its members 
than the first group. 
These observations corroborate previous measures. 

\section{Discussion}
\label{sec-discussion}
Expectedly, methods characterized by different attributes
yield distinct results as we have seen in \Cref{sec-experiments}.
The branch decomposition-based methods are theoretically
proven to be prone to instabilities \changed{when computing} the edit mapping.
Hence, they are expected to 
\changed{differ significantly} 
from
the methods aimed at alleviating the instability. 
The statement holds for multiple cases as shown in previous works 
\cite{saikia14a, wetzels2022branch,wetzels2022path}. 
However, these improvements can come at the cost of 
excluding the saddle points (see \Cref{sec-verticalHorizontal})
or significant computational overhead. 
In the experiment detailed in \Cref{sec-stableNoise}, 
the branch decomposition-independent methods
were unable to produce any result in a reasonable time 
when setting $\tau = 20\%$, corresponds to merge trees with around $470$ nodes.  
In fact, creating a method that handles 
every instability can be intractable. 
The look-ahead parameter of the path mapping methods \cite{wetzels2025stable}
makes a trade-off between practical feasibility and theoretical stability.
This method constantly shows higher performance in tracking results.
However, since it introduced a heuristic that carefully handles special cases,
the question arises whether similar improvements could be
achieved in other methods through more optimized designs.

The use of merge trees for feature tracking often 
considers each time step as an independent domain
and establishes connections solely between the nodes of the trees.
This approach often excludes the temporal \changed{and physical} information,
which can lead to suboptimal tracking results.
For instance, consider a feature that is insignificant 
initially but becomes increasingly prominent over time. 
Persistent-based information independently extracted 
from each merge tree may fail to capture such a trend
and may not track the feature at all. 
Consequently, it remains an open question whether 
and how \changed{additional} information can be effectively incorporated   
into merge tree-based feature tracking.

The similarity between methods can vary by data set.
A method can appear more similar to a second one 
than a third one in this data set,
but this pattern may not replicate in a different data set. 
While recent methods for comparing and feature tracking 
using merge trees have each addressed the shortcomings of their predecessors,
\changed{the suitability of the data sets 
generated by frameworks such as
that proposed by Nilsson et al. \cite{nilsson2022towards}
in covering all relevant topological cases needs to be tested.}
\changed{In addition, since existing frameworks focus 
on general tracking evaluation,
there is no standardized framework specifically designed 
to account for the properties of merge trees.
}
The comparison approaches presented in this paper,
as trivial as they are, show their ability to distinguish
the tracked results. 
This argument highlights the need for the development of 
robust evaluation frameworks tailored to this context.

\section{Conclusion}
\label{sec-conclusion}

We compared four different methods for 
feature tracking using merge trees.
We analyzed their behaviors on analytical data sets 
specifically designed to investigate the
problems of vertical and horizontal instability. 
\changed{A data set with verifiable topological features
is created, 
enabling us to assess 
the stability of the selected methods.}
Our results showed  that despite the discrepancies, 
those methods are relatively stable under low levels of noise.
Additionally, we conducted experiments on real-world data sets 
from the fields of fluid dynamics and medical imaging. 
We evaluated the individual performance of each method
on these data sets as well as pairwise similarities
between the methods to measure their agreements
in more practical scenarios.
Our results showed that the differences among 
the methods arise not only from their theoretical foundations,
but also from underlying characteristics of the data sets, 
which play crucial roles. 

In this paper, we focused on four methods that 
are largely based on the edit distance. 
While differences among them are expected,
future works can include quantifying how distinct
this class of methods is compared to 
other tailored merge tree-based feature tracking methods
\changed{and more advanced methods that are considered as standards for feature tracking}.
Given the \changed{scarcity} of ground truth data for merge tree tracking,
our approach can be seen as \changed{one of the first steps} toward addressing the challenge
of finding meaningful comparison metrics.
We wish to further investigate this problem.

Despite highlighting the
variability among methods
our comparisons did not assess the quality of the 
tracking results, 
that is whether the tracking of a feature is meaningful. 
Developing a flexible and robust system
for visually comparing and qualitatively assessing  
the tracking results by domain experts could be helpful in this regard.
Moreover, constructing a theoretical metric
operating on the space of edit distances
between merge trees, 
or more broadly, 
on the space of distances between merge trees 
or the space of tracking paths produced 
by different methods, remains an open and important question.

%
%
%
%
%
%

\acknowledgments{
This work was supported through grants from The Swedish Research Council (Vetenskapsrådet, project 2020-05461), 
and the Swedish e-Science Research Centre (SeRC).
We would like to thank Joel Kronborg and Johan Hoffman for providing us with \dataset{HeartBeat3D} data set. 
We would like to thank the anonymous reviewers for their valuable feedback.
}

\bibliographystyle{abbrv-doi}

\bibliography{Literature}

\begin{thebibliography}{10}

\bibitem{bauer02}
D.~Bauer and R.~Peikert.
\newblock Vortex tracking in scale space.
\newblock In {\em Data Visualization 2002. Proc.\ VisSym 02}, pp. 233--240,
  2002. doi: {{%
10\hspace{.1pt}\discretionary{.}{%
}{.}\hspace{.4pt}5555\discretionary{/}{%
}{/}509740\hspace{.1pt}\discretionary{.}{%
}{.}\hspace{.4pt}509779}}


\bibitem{beketayev2014}
K.~Beketayev, D.~Yeliussizov, D.~Morozov, G.~H. Weber, and B.~Hamann.
\newblock Measuring the distance between merge trees.
\newblock In P.-T. Bremer, I.~Hotz, V.~Pascucci, and R.~Peikert, eds., {\em
  Topological Methods in Data Analysis and Visualization III}, pp. 151--165.
  Springer International Publishing, Cham, 2014.

\bibitem{chen2013topology}
F.~Chen, H.~Obermaier, H.~Hagen, B.~Hamann, J.~Tierny, and V.~Pascucci.
\newblock Topology analysis of time-dependent multi-fluid data using the reeb
  graph.
\newblock {\em Computer Aided Geometric Design}, 30(6):557--566, 2013.

\bibitem{das2024time}
S.~Das, R.~Sridharamurthy, and V.~Natarajan.
\newblock Time-varying extremum graphs.
\newblock In {\em Comput. Graph. Forum}, vol.~43, p. e15162. Wiley Online
  Library, 2024.

\bibitem{edelsbrunner2004time}
H.~Edelsbrunner, J.~Harer, A.~Mascarenhas, and V.~Pascucci.
\newblock Time-varying reeb graphs for continuous space-time data.
\newblock In {\em Proceedings of the twentieth annual symposium on
  Computational geometry}, pp. 366--372, 2004.

\bibitem{edelsbrunner02}
H.~Edelsbrunner, D.~Letscher, and A.~Zomorodian.
\newblock Topological persistence and simplification.
\newblock {\em Discrete and Computational Geometry}, 28(4):511 -- 533, 2002.

\bibitem{forman98}
R.~Forman.
\newblock {M}orse theory for cell-complexes.
\newblock {\em Advances in Mathematics}, 134(1):90--145, 1998.

\bibitem{garth04b}
C.~Garth, X.~Tricoche, and G.~Scheuermann.
\newblock Tracking of vector field singularities in unstructured {3D}
  time-dependent datasets.
\newblock In {\em Proc.\ IEEE Visualization}, pp. 329--336, 2004.

\bibitem{gasparovic2025intrinsic}
E.~Gasparovic, E.~Munch, S.~Oudot, K.~Turner, B.~Wang, and Y.~Wang.
\newblock Intrinsic interleaving distance for merge trees.
\newblock {\em La Matematica}, 4(1):40--65, 2025.

\bibitem{germer11a}
T.~Germer, M.~Otto, R.~Peikert, and H.~Theisel.
\newblock Lagrangian coherent structures with guaranteed material separation.
\newblock {\em Comput. Graph. Forum (Proc.\ EuroVis)}, 30(3):761--770, 2011.

\bibitem{Guenther15SciVis}
T.~G{\"u}nther, M.~Schulze, and H.~Theisel.
\newblock Rotation invariant vortices for flow visualization.
\newblock {\em IEEE Trans. Vis. Comput. Graph. (IEEE Visualization)},
  22(1):817--826, 2016. doi: {{%
10\hspace{.1pt}\discretionary{.}{%
}{.}\hspace{.4pt}1109\discretionary{/}{%
}{/}TVCG\hspace{.1pt}\discretionary{.}{%
}{.}\hspace{.4pt}2015\hspace{.1pt}\discretionary{.}{%
}{.}\hspace{.4pt}2467200}}


\bibitem{koeppweinkauf2019}
W.~K{\"o}pp and T.~Weinkauf.
\newblock Temporal treemaps: Static visualization of evolving trees.
\newblock {\em {IEEE} Trans. Vis. Comput. Graph.}, 25(1):534--543, 2019. doi:
  {{%
10\hspace{.1pt}\discretionary{.}{%
}{.}\hspace{.4pt}1109\discretionary{/}{%
}{/}TVCG\hspace{.1pt}\discretionary{.}{%
}{.}\hspace{.4pt}2018\hspace{.1pt}\discretionary{.}{%
}{.}\hspace{.4pt}2865265}}


\bibitem{koepp22}
W.~Köpp and T.~Weinkauf.
\newblock Temporal merge tree maps: A topology-based static visualization for
  temporal scalar data.
\newblock {\em {IEEE} Trans. Vis. Comput. Graph.}, 29(1):1--11, Oct. 2022. doi:
  {{%
10\hspace{.1pt}\discretionary{.}{%
}{.}\hspace{.4pt}1109\discretionary{/}{%
}{/}TVCG\hspace{.1pt}\discretionary{.}{%
}{.}\hspace{.4pt}2022\hspace{.1pt}\discretionary{.}{%
}{.}\hspace{.4pt}3209387}}


\bibitem{larsson2017heart}
D.~Larsson, J.~H. Sp\"uhler, S.~Petersson, T.~Nordenfur, M.~Colarieti-Tosti,
  J.~Hoffman, R.~Winter, and M.~Larsson.
\newblock Patient-specific left ventricular flow simulations from transthoracic
  echocardiography: Robustness evaluation and validation against ultrasound
  doppler and magnetic resonance imaging.
\newblock {\em IEEE Trans. Medical Imaging}, 36(11):2261--2275, 2017. doi: {{%
10\hspace{.1pt}\discretionary{.}{%
}{.}\hspace{.4pt}1109\discretionary{/}{%
}{/}TMI\hspace{.1pt}\discretionary{.}{%
}{.}\hspace{.4pt}2017\hspace{.1pt}\discretionary{.}{%
}{.}\hspace{.4pt}2718218}}


\bibitem{li2025ot}
M.~Li, X.~Yan, L.~Yan, T.~Needham, and B.~Wang.
\newblock Flexible and probabilistic topology tracking with partial optimal
  transport.
\newblock {\em IEEE Tran. Vis. Comput. Graph.}, pp. 1--18, 2025. doi: {{%
10\hspace{.1pt}\discretionary{.}{%
}{.}\hspace{.4pt}1109\discretionary{/}{%
}{/}TVCG\hspace{.1pt}\discretionary{.}{%
}{.}\hspace{.4pt}2025\hspace{.1pt}\discretionary{.}{%
}{.}\hspace{.4pt}3561300}}


\bibitem{lohfinketal2021}
A.-P. Lohfink, F.~Gartzky, F.~Wetzels, L.~Vollmer, and C.~Garth.
\newblock {Time-Varying Fuzzy Contour Trees}.
\newblock In {\em 2021 IEEE Visualization Conference (VIS)}, pp. 86--90, 2021.
  doi: {{%
10\hspace{.1pt}\discretionary{.}{%
}{.}\hspace{.4pt}1109\discretionary{/}{%
}{/}VIS49827\hspace{.1pt}\discretionary{.}{%
}{.}\hspace{.4pt}2021\hspace{.1pt}\discretionary{.}{%
}{.}\hspace{.4pt}9623286}}


\bibitem{lohfinketal2020}
A.-P. Lohfink, F.~Wetzels, J.~Lukasczyk, G.~H. Weber, and C.~Garth.
\newblock Fuzzy contour trees: Alignment and joint layout of multiple contour
  trees.
\newblock {\em Comput. Graph. Forum (Proc. EuroVis)}, 39(3):343--355, 2020.
  doi: {{%
10\hspace{.1pt}\discretionary{.}{%
}{.}\hspace{.4pt}1111\discretionary{/}{%
}{/}cgf\hspace{.1pt}\discretionary{.}{%
}{.}\hspace{.4pt}13985}}


\bibitem{lukasczyketal2020}
J.~Lukasczyk, C.~Garth, G.~H. Weber, T.~Biedert, R.~Maciejewski, and H.~Leitte.
\newblock Dynamic nested tracking graphs.
\newblock {\em IEEE Transactions on Visualization \& Computer Graphics (Proc.
  IEEE VIS)}, 26(1):249--258, 2020. doi: {{%
10\hspace{.1pt}\discretionary{.}{%
}{.}\hspace{.4pt}1109\discretionary{/}{%
}{/}TVCG\hspace{.1pt}\discretionary{.}{%
}{.}\hspace{.4pt}2019\hspace{.1pt}\discretionary{.}{%
}{.}\hspace{.4pt}2934368}}


\bibitem{lukasczyketal2017}
J.~Lukasczyk, G.~Weber, R.~Maciejewski, C.~Garth, and H.~Leitte.
\newblock Nested tracking graphs.
\newblock {\em Comput. Graph. Forum}, 36(3):643--667, 2017. doi: {{%
10\hspace{.1pt}\discretionary{.}{%
}{.}\hspace{.4pt}1111\discretionary{/}{%
}{/}cgf\hspace{.1pt}\discretionary{.}{%
}{.}\hspace{.4pt}13164}}


\bibitem{lyu2025lsh}
W.~Lyu, R.~Sridharamurthy, J.~M. Phillips, and B.~Wang.
\newblock Fast comparative analysis of merge trees using locality sensitive
  hashing.
\newblock {\em IEEE Trans. Vis. Comput. Graph.}, 31(1):141--151, 2025. doi: {{%
10\hspace{.1pt}\discretionary{.}{%
}{.}\hspace{.4pt}1109\discretionary{/}{%
}{/}TVCG\hspace{.1pt}\discretionary{.}{%
}{.}\hspace{.4pt}2024\hspace{.1pt}\discretionary{.}{%
}{.}\hspace{.4pt}3456383}}


\bibitem{narayanan2015distance}
V.~Narayanan, D.~M. Thomas, and V.~Natarajan.
\newblock Distance between extremum graphs.
\newblock In {\em 2015 IEEE Pacific Visualization Symposium (PacificVis)}, pp.
  263--270. IEEE, 2015.

\bibitem{neu13}
U.~Neu, M.~G. Akperov, N.~Bellenbaum, R.~Benestad, R.~Blender, R.~Caballero,
  A.~Cocozza, H.~F. Dacre, Y.~Feng, K.~Fraedrich, et~al.
\newblock Imilast: a community effort to intercompare extratropical cyclone
  detection and tracking algorithms.
\newblock {\em Bulletin of the American Meteorological Society},
  94(4):529--547, 2013.

\bibitem{nilsson2022towards}
E.~Nilsson, J.~Lukasczyk, T.~B. Masood, C.~Garth, and I.~Hotz.
\newblock Towards benchmark data generation for feature tracking in scalar
  fields.
\newblock In {\em 2022 Topological Data Analysis and Visualization
  (TopoInVis)}, pp. 103--112. IEEE, 2022.

\bibitem{oesterling15}
P.~Oesterling, C.~Heine, G.~H. Weber, D.~Morozov, and G.~Scheuermann.
\newblock Computing and visualizing time-varying merge trees for
  high-dimensional data.
\newblock In H.~Carr, C.~Garth, and T.~Weinkauf, eds., {\em Topological Methods
  in Data Analysis and Visualization IV}, pp. 87--101. Springer International
  Publishing, 2017. doi: {{%
10\hspace{.1pt}\discretionary{.}{%
}{.}\hspace{.4pt}1007\discretionary{/}{%
}{/}978\discretionary{%
}{-}{-}3\discretionary{%
}{-}{-}319\discretionary{%
}{-}{-}44684\discretionary{%
}{-}{-}4\_5}}


\bibitem{pont22}
M.~Pont, J.~Vidal, J.~Delon, and J.~Tierny.
\newblock Wasserstein distances, geodesics and barycenters of merge trees.
\newblock {\em {IEEE} Trans. Vis. Comput. Graph.}, 28(1):291--301, 2022. doi:
  {{%
10\hspace{.1pt}\discretionary{.}{%
}{.}\hspace{.4pt}1109\discretionary{/}{%
}{/}tvcg\hspace{.1pt}\discretionary{.}{%
}{.}\hspace{.4pt}2021\hspace{.1pt}\discretionary{.}{%
}{.}\hspace{.4pt}3114839}}


\bibitem{qin2025fast}
Y.~Qin, B.~T. Fasy, C.~Wenk, and B.~Summa.
\newblock Rapid and precise topological comparison with merge tree neural
  networks.
\newblock {\em IEEE Trans. Vis. Comput. Graph.}, 31(1):1322--1332, 2025. doi:
  {{%
10\hspace{.1pt}\discretionary{.}{%
}{.}\hspace{.4pt}1109\discretionary{/}{%
}{/}TVCG\hspace{.1pt}\discretionary{.}{%
}{.}\hspace{.4pt}2024\hspace{.1pt}\discretionary{.}{%
}{.}\hspace{.4pt}3456395}}


\bibitem{saikia14a}
H.~Saikia, H.-P. Seidel, and T.~Weinkauf.
\newblock Extended branch decomposition graphs: Structural comparison of scalar
  data.
\newblock {\em Comput. Graph. Forum}, 33(3):41--50, June 2014.

\bibitem{saikia16a}
H.~Saikia and T.~Weinkauf.
\newblock Global feature tracking and similarity estimation in time-dependent
  scalar fields.
\newblock {\em Computer Graphics Forum}, 36(3):1--11, June 2017. doi: {{%
10\hspace{.1pt}\discretionary{.}{%
}{.}\hspace{.4pt}1111\discretionary{/}{%
}{/}cgf\hspace{.1pt}\discretionary{.}{%
}{.}\hspace{.4pt}13163}}


\bibitem{silverwang1997}
D.~Silver and X.~Wang.
\newblock Tracking and visualizing turbulent {3D} features.
\newblock {\em IEEE Trans. Vis. Comput. Graph.}, 3(2):129--141, 1997. doi: {{%
10\hspace{.1pt}\discretionary{.}{%
}{.}\hspace{.4pt}1109\discretionary{/}{%
}{/}2945\hspace{.1pt}\discretionary{.}{%
}{.}\hspace{.4pt}597796}}


\bibitem{silverwang1998}
D.~Silver and X.~Wang.
\newblock Tracking scalar features in unstructured datasets.
\newblock In D.~S. Ebert, H.~E. Rushmeier, and H.~Hagen, eds., {\em
  Visualization 9´8, Proceedings, October 18-23, 1998, Research Triangle Park,
  North Carolina, USA.}, pp. 79--86. IEEE Computer Society and ACM, 1998. doi:
  {{%
10\hspace{.1pt}\discretionary{.}{%
}{.}\hspace{.4pt}1109\discretionary{/}{%
}{/}VISUAL\hspace{.1pt}\discretionary{.}{%
}{.}\hspace{.4pt}1998\hspace{.1pt}\discretionary{.}{%
}{.}\hspace{.4pt}745288}}


\bibitem{sohn06}
B.~S. Sohn and C.~Bajaj.
\newblock Time-varying contour topology.
\newblock {\em IEEE TVCG}, 12(1):14--25, Jan 2006. doi: {{%
10\hspace{.1pt}\discretionary{.}{%
}{.}\hspace{.4pt}1109\discretionary{/}{%
}{/}TVCG\hspace{.1pt}\discretionary{.}{%
}{.}\hspace{.4pt}2006\hspace{.1pt}\discretionary{.}{%
}{.}\hspace{.4pt}16}}


\bibitem{soler2019ranking}
M.~Soler, M.~Petitfrere, G.~Darche, M.~Plainchault, B.~Conche, and J.~Tierny.
\newblock Ranking viscous finger simulations to an acquired ground truth with
  topology-aware matchings.
\newblock In {\em 2019 IEEE 9th Symposium on Large Data Analysis and
  Visualization (LDAV)}, pp. 62--72. IEEE, 2019.

\bibitem{soleretal2018}
M.~Soler, M.~Plainchault, B.~Conche, and J.~Tierny.
\newblock Lifted wasserstein matcher for fast and robust topology tracking.
\newblock In {\em 2018 IEEE 8th Symposium on Large Data Analysis and
  Visualization (LDAV)}, pp. 23--33, 2018. doi: {{%
10\hspace{.1pt}\discretionary{.}{%
}{.}\hspace{.4pt}1109\discretionary{/}{%
}{/}LDAV\hspace{.1pt}\discretionary{.}{%
}{.}\hspace{.4pt}2018\hspace{.1pt}\discretionary{.}{%
}{.}\hspace{.4pt}8739196}}


\bibitem{sridharamurthy2018edit}
R.~Sridharamurthy, T.~B. Masood, A.~Kamakshidasan, and V.~Natarajan.
\newblock Edit distance between merge trees.
\newblock {\em {IEEE} Trans. Vis. Comput. Graph.}, 26(3):1518--1531, 2018.

\bibitem{tai1979}
K.-C. Tai.
\newblock The tree-to-tree correction problem.
\newblock {\em J. ACM}, 26(3):422–433, July 1979. doi: {{%
10\hspace{.1pt}\discretionary{.}{%
}{.}\hspace{.4pt}1145\discretionary{/}{%
}{/}322139\hspace{.1pt}\discretionary{.}{%
}{.}\hspace{.4pt}322143}}


\bibitem{theisel05b}
H.~Theisel, J.~Sahner, T.~Weinkauf, H.-C. Hege, and H.-P. Seidel.
\newblock {Extraction of Parallel Vector Surfaces in {3D} Time-Dependent Fields
  and Application to Vortex Core Line Tracking}.
\newblock In {\em Proc.\ IEEE Visualization 2005}, pp. 631--638, 2005. doi: {{%
10\hspace{.1pt}\discretionary{.}{%
}{.}\hspace{.4pt}1109\discretionary{/}{%
}{/}VISUAL\hspace{.1pt}\discretionary{.}{%
}{.}\hspace{.4pt}2005\hspace{.1pt}\discretionary{.}{%
}{.}\hspace{.4pt}1532851}}


\bibitem{theisel03b}
H.~Theisel and H.-P. Seidel.
\newblock {Feature Flow Fields}.
\newblock In {\em Data Visualization 2003. Proc.\ VisSym 03}, pp. 141--148,
  2003. doi: {{%
10\hspace{.1pt}\discretionary{.}{%
}{.}\hspace{.4pt}5555\discretionary{/}{%
}{/}769922\hspace{.1pt}\discretionary{.}{%
}{.}\hspace{.4pt}769938}}


\bibitem{thomas11}
D.~M. Thomas and V.~Natarajan.
\newblock Symmetry in scalar field topology.
\newblock {\em IEEE TVCG}, 17(12):2035--2044, 2011.

\bibitem{tiernyetal2018}
J.~Tierny, G.~Favelier, J.~A. Levine, C.~Gueunet, and M.~Michaux.
\newblock The topology toolkit.
\newblock {\em {IEEE} Trans. Vis. Comput. Graph.}, 24(1):832--842, 2018. doi:
  {{%
10\hspace{.1pt}\discretionary{.}{%
}{.}\hspace{.4pt}1109\discretionary{/}{%
}{/}TVCG\hspace{.1pt}\discretionary{.}{%
}{.}\hspace{.4pt}2017\hspace{.1pt}\discretionary{.}{%
}{.}\hspace{.4pt}2743938}}


\bibitem{tricoche02}
X.~Tricoche, T.~Wischgoll, G.~Scheuermann, and H.~Hagen.
\newblock Topology tracking for the visualization of time-dependent
  two-dimensional flows.
\newblock {\em Computers \& Graphics}, 26:249--257, 2002.

\bibitem{weber11}
G.~Weber, P.-T. Bremer, M.~Day, J.~Bell, and V.~Pascucci.
\newblock Feature tracking using reeb graphs.
\newblock In V.~Pascucci, X.~Tricoche, H.~Hagen, and J.~Tierny, eds., {\em
  Topological Methods in Data Analysis and Visualization: Theory, Algorithms,
  and Applications}, pp. 241--253. Springer Berlin Heidelberg, Berlin,
  Heidelberg, 2011. doi: {{%
10\hspace{.1pt}\discretionary{.}{%
}{.}\hspace{.4pt}1007\discretionary{/}{%
}{/}978\discretionary{%
}{-}{-}3\discretionary{%
}{-}{-}642\discretionary{%
}{-}{-}15014\discretionary{%
}{-}{-}2\_20}}


\bibitem{weinkauf07c}
T.~Weinkauf, J.~Sahner, H.~Theisel, and H.-C. Hege.
\newblock Cores of swirling particle motion in unsteady flows.
\newblock {\em IEEE Trans. Vis. Comput. Graph. (Proceedings Visualization
  2007)}, 13(6):1759--1766, November -- December 2007.

\bibitem{weinkauf10c}
T.~Weinkauf and H.~Theisel.
\newblock Streak lines as tangent curves of a derived vector field.
\newblock {\em IEEE Trans. Vis. Comput. Graph. (Proc. IEEE Visualization)},
  16(6):1225--1234, November - December 2010.

\bibitem{weinkauf10a}
T.~Weinkauf, H.~Theisel, A.~V. Gelder, and A.~Pang.
\newblock Stable feature flow fields.
\newblock {\em IEEE Trans. Vis. Comput. Graph.}, 2010.
\newblock accepted.

\bibitem{wetzels2023stable}
F.~Wetzels, M.~Anders, and C.~Garth.
\newblock Taming horizontal instability in merge trees: On the computation of a
  comprehensive deformation-based edit distance.
\newblock In {\em 2023 Topological Data Analysis and Visualization
  (TopoInVis)}, pp. 82--92, 2023. doi: {{%
10\hspace{.1pt}\discretionary{.}{%
}{.}\hspace{.4pt}1109\discretionary{/}{%
}{/}TopoInVis60193\hspace{.1pt}\discretionary{.}{%
}{.}\hspace{.4pt}2023\hspace{.1pt}\discretionary{.}{%
}{.}\hspace{.4pt}00015}}


\bibitem{wetzels2022path}
F.~Wetzels and C.~Garth.
\newblock A deformation-based edit distance for merge trees.
\newblock In {\em 2022 Topological Data Analysis and Visualization
  (TopoInVis)}, pp. 29--38, 2022. doi: {{%
10\hspace{.1pt}\discretionary{.}{%
}{.}\hspace{.4pt}1109\discretionary{/}{%
}{/}TopoInVis57755\hspace{.1pt}\discretionary{.}{%
}{.}\hspace{.4pt}2022\hspace{.1pt}\discretionary{.}{%
}{.}\hspace{.4pt}00010}}


\bibitem{wetzels2022branch}
F.~Wetzels, H.~Leitte, and C.~Garth.
\newblock Branch decomposition-independent edit distances for merge trees.
\newblock In {\em Comput. Graph. Forum}, vol.~41, pp. 367--378. Wiley Online
  Library, 2022.

\bibitem{wetzels2025stable}
F.~Wetzels, H.~Leitte, and C.~Garth.
\newblock Accelerating computation of stable merge tree edit distances using
  parameterized heuristics.
\newblock {\em IEEE Trans. Vis. Comput. Graph.}, 31(6):3706--3718, 2025. doi:
  {{%
10\hspace{.1pt}\discretionary{.}{%
}{.}\hspace{.4pt}1109\discretionary{/}{%
}{/}TVCG\hspace{.1pt}\discretionary{.}{%
}{.}\hspace{.4pt}2025\hspace{.1pt}\discretionary{.}{%
}{.}\hspace{.4pt}3567120}}


\bibitem{yan2022geometry}
L.~Yan, T.~B. Masood, F.~Rasheed, I.~Hotz, and B.~Wang.
\newblock Geometry-aware merge tree comparisons for time-varying data with
  interleaving distances.
\newblock {\em IEEE Trans. Vis. Comput. Graph.}, 29(8):3489--3506, 2022.

\bibitem{yan2021scalar}
L.~Yan, T.~B. Masood, R.~Sridharamurthy, F.~Rasheed, V.~Natarajan, I.~Hotz, and
  B.~Wang.
\newblock Scalar field comparison with topological descriptors: Properties and
  applications for scientific visualization.
\newblock In {\em Comput. Graph. Forum}, vol.~40, pp. 599--633. Wiley Online
  Library, 2021.

\bibitem{zhang1996constrained}
K.~Zhang.
\newblock A constrained edit distance between unordered labeled trees.
\newblock {\em Algorithmica}, 15(3):205--222, Mar 1996. doi: {{%
10\hspace{.1pt}\discretionary{.}{%
}{.}\hspace{.4pt}1007\discretionary{/}{%
}{/}BF01975866}}


\bibitem{zhang1992}
K.~Zhang, R.~Statman, and D.~Shasha.
\newblock On the editing distance between unordered labeled trees.
\newblock {\em Information Processing Letters}, 42(3):133--139, 1992. doi: {{%
10\hspace{.1pt}\discretionary{.}{%
}{.}\hspace{.4pt}1016\discretionary{/}{%
}{/}0020\discretionary{%
}{-}{-}0190\discretionary{%
}{(}{(}92\discretionary{)}{%
}{)}90136\discretionary{%
}{-}{-}J}}


\end{thebibliography}


\begin{thebibliography}{10}

\bibitem{BaezaRojo19SciVisa}
I.~Baeza~Rojo and T.~G{\"u}nther.
\newblock Vector field topology of time-dependent flows in a steady reference
  frame.
\newblock {\em {IEEE} Trans. Vis. Comput. Graph.}, 2019.

\bibitem{germer11a}
T.~Germer, M.~Otto, R.~Peikert, and H.~Theisel.
\newblock Lagrangian coherent structures with guaranteed material separation.
\newblock {\em Comput. Graph. Forum (Proc.\ EuroVis)}, 30(3):761--770, 2011.

\bibitem{Guenther17}
T.~G{\"u}nther, M.~Gross, and H.~Theisel.
\newblock Generic objective vortices for flow visualization.
\newblock {\em {ACM} Trans. Graph.}, 36(4):141:1--141:11, 2017.

\bibitem{ERA5data}
H.~Hersbach, B.~Bell, P.~Berrisford, G.~Biavati, A.~Horányi,
  J.~Muñoz~Sabater, J.~Nicolas, C.~Peubey, R.~Radu, I.~Rozum, D.~Schepers,
  A.~Simmons, C.~Soci, D.~Dee, and J.-N. Thépaut.
\newblock {ERA5 hourly data on single levels from 1979 to present}.
\newblock Copernicus Climate Change Service (C3S) Climate Data Store (CDS),
  2018.
\newblock Accessed on 10-Mar-2022. doi: {{%
10\hspace{.1pt}\discretionary{.}{%
}{.}\hspace{.4pt}24381\discretionary{/}{%
}{/}cds\hspace{.1pt}\discretionary{.}{%
}{.}\hspace{.4pt}adbb2d47}}


\bibitem{koepp22}
W.~Köpp and T.~Weinkauf.
\newblock Temporal merge tree maps: A topology-based static visualization for
  temporal scalar data.
\newblock {\em {IEEE} Trans. Vis. Comput. Graph.}, 29(1):1--11, Oct. 2022. doi:
  {{%
10\hspace{.1pt}\discretionary{.}{%
}{.}\hspace{.4pt}1109\discretionary{/}{%
}{/}TVCG\hspace{.1pt}\discretionary{.}{%
}{.}\hspace{.4pt}2022\hspace{.1pt}\discretionary{.}{%
}{.}\hspace{.4pt}3209387}}


\bibitem{pont22}
M.~Pont, J.~Vidal, J.~Delon, and J.~Tierny.
\newblock Wasserstein distances, geodesics and barycenters of merge trees.
\newblock {\em {IEEE} Trans. Vis. Comput. Graph.}, 28(1):291--301, 2022. doi:
  {{%
10\hspace{.1pt}\discretionary{.}{%
}{.}\hspace{.4pt}1109\discretionary{/}{%
}{/}tvcg\hspace{.1pt}\discretionary{.}{%
}{.}\hspace{.4pt}2021\hspace{.1pt}\discretionary{.}{%
}{.}\hspace{.4pt}3114839}}


\bibitem{PopinetEtAl2004_Tangaroa}
S.~Popinet, M.~Smith, and C.~Stevens.
\newblock {Experimental and Numerical Study of the Turbulence Characteristics
  of Airflow around a Research Vessel}.
\newblock {\em J. Atmos. Ocean. Technol.}, 21(10):1575--1589, oct 2004. doi:
  {{%
10\hspace{.1pt}\discretionary{.}{%
}{.}\hspace{.4pt}1175\discretionary{/}{%
}{/}1520\discretionary{%
}{-}{-}0426\discretionary{%
}{(}{(}2004\discretionary{)}{%
}{)}021{\textless}1575\discretionary{:}{%
}{:}eansot{\textgreater}2\hspace{.1pt}\discretionary{.}{%
}{.}\hspace{.4pt}0\hspace{.1pt}\discretionary{.}{%
}{.}\hspace{.4pt}co\discretionary{;}{%
}{;}2}}


\bibitem{sridharamurthy2018edit}
R.~Sridharamurthy, T.~B. Masood, A.~Kamakshidasan, and V.~Natarajan.
\newblock Edit distance between merge trees.
\newblock {\em {IEEE} Trans. Vis. Comput. Graph.}, 26(3):1518--1531, 2018.

\bibitem{wetzels2022path}
F.~Wetzels and C.~Garth.
\newblock A deformation-based edit distance for merge trees.
\newblock In {\em 2022 Topological Data Analysis and Visualization
  (TopoInVis)}, pp. 29--38, 2022. doi: {{%
10\hspace{.1pt}\discretionary{.}{%
}{.}\hspace{.4pt}1109\discretionary{/}{%
}{/}TopoInVis57755\hspace{.1pt}\discretionary{.}{%
}{.}\hspace{.4pt}2022\hspace{.1pt}\discretionary{.}{%
}{.}\hspace{.4pt}00010}}


\bibitem{wetzels2022branch}
F.~Wetzels, H.~Leitte, and C.~Garth.
\newblock Branch decomposition-independent edit distances for merge trees.
\newblock In {\em Comput. Graph. Forum}, vol.~41, pp. 367--378. Wiley Online
  Library, 2022.

\bibitem{whalen2008competition}
D.~Whalen and M.~Norman.
\newblock Competition data set and description.
\newblock {\em 2008 IEEE Visualization Design Contest}, 2008.

\end{thebibliography}
\end{document}